\documentclass[twocolumn]{aastex631}

\usepackage{xcolor}
\usepackage{ulem}
\usepackage{soul}
\begin{document}

\title{Kpc-Scale Neutral Iron K$\alpha$ Emission in the Starburst-AGN NGC 4945: a Relic AGN Outflow?

}

\author[0009-0008-4232-486X]{Kimberly A. Weaver}
\affiliation{X-ray Astrophysics Laboratory, NASA Goddard Space Flight Center, Code 662, Greenbelt, MD 20771, USA}

\author[0000-0003-1051-6564]{Jenna M. Cann}
\altaffiliation{NASA Postdoctoral Program Fellow}
\affiliation{X-ray Astrophysics Laboratory, NASA Goddard Space Flight Center, Code 662, Greenbelt, MD 20771, USA}
\affiliation{Oak Ridge Associated Universities, NASA NPP Program, Oak Ridge, TN 37831, USA}

\author{Lynne Valencic}
\affiliation{X-ray Astrophysics Laboratory, NASA Goddard Space Flight Center, Code 662, Greenbelt, MD 20771, USA}
\affiliation{Johns Hopkins University}

\author[0000-0001-8640-8522]{Ryan W. Pfeifle}
\altaffiliation{NASA Postdoctoral Program Fellow}
\affiliation{X-ray Astrophysics Laboratory, NASA Goddard Space Flight Center, Code 662, Greenbelt, MD 20771, USA}
\affiliation{Oak Ridge Associated Universities, NASA NPP Program, Oak Ridge, TN 37831, USA}

\author{K. D. Kuntz}
\affiliation{X-ray Astrophysics Laboratory, NASA Goddard Space Flight Center, Code 662, Greenbelt, MD 20771, USA}
\affiliation{Johns Hopkins University}

\author[0000-0001-8873-7450]{Joel F. Campbell}
\affiliation{NASA Langley Research Center}

\author{Kimberly Engle}
\affiliation{X-ray Astrophysics Laboratory, NASA Goddard Space Flight Center, Code 662, Greenbelt, MD 20771, USA}
\affiliation{ADNET Systems, Inc., Lanham, MD 20706, USA}

\author[0000-0002-1359-1626]{Ryan Tanner}
\affiliation{X-ray Astrophysics Laboratory, NASA Goddard Space Flight Center, Code 662, Greenbelt, MD 20771, USA}
\affiliation{Catholic University, Washington DC 20064, USA}

\author{Edmund Hodges-Kluck}
\affiliation{X-ray Astrophysics Laboratory, NASA Goddard Space Flight Center, Code 662, Greenbelt, MD 20771, USA}

\author[0000-0003-1172-5018]{Isabella Carlton}
\affiliation{Southeastern Universities Research Association, Washington DC 20005, USA}
\affiliation{X-ray Astrophysics Laboratory, NASA Goddard Space Flight Center, Code 662, Greenbelt, MD 20771, USA}
\affiliation{Center for Research and Exploration in Space Science and Technology, NASA/GSFC, Greenbelt, MD 20771}

\author{Miranda McCarthy}
\affiliation{Southeastern Universities Research Association, Washington DC 20005, USA}
\affiliation{X-ray Astrophysics Laboratory, NASA Goddard Space Flight Center, Code 662, Greenbelt, MD 20771, USA}
\affiliation{Center for Research and Exploration in Space Science and Technology, NASA/GSFC, Greenbelt, MD 20771}



\begin{abstract}

NGC 4945 contains a well-known heavily obscured active galactic nucleus (AGN) at its core, with prior reports of strong nuclear and off-nuclear neutral Fe K$\alpha$ emission due to the AGN activity.\ We report the discovery of very extended Fe K$\alpha$ emission with the XMM-Newton EPIC pn in a $\sim5$\,kpc by $\sim10$\,kpc region that is misaligned with the plane of the inclined optical galaxy disk by $\sim60$ degrees in projection.\ After a careful consideration of the crowded center of the galaxy and numerous unresolved hard X-ray sources present, we estimate that $\sim15\%$ of the Fe K$\alpha$ is extended on kpc-sized scales.\ The overall size and misalignment of the region follows an unusual pattern of radio polarization that is not typical of starbursts or normal disk galaxies but has been interpreted as possibly due to AGN activity.\ We suggest that the extended Fe K$\alpha$ emission arose from a period of AGN eruption several million years ago - a relic of a past AGN ejection episode.


\end{abstract}



\section{Introduction} \label{sec:intro}

Astronomers are certain that the evolution of galaxies is regulated by galactic outflows, but the specific driving mechanisms for these outflows are debatable.\ In the nearby universe, starburst-generated outflows are common \citep[e.g.][]{1990ApJS...74..833H}.\ 
It also seems clear that the most luminous nearby active galactic nuclei (AGN) can drive powerful outflows.\ Much less evidence exists for AGN-generated outflows and feedback in lower luminosity, more typical, Seyfert galaxies \citep{Heckman_2014}.\ 
On the other hand, AGN are compact and highly variable, and their luminosities can rise and decay quickly compared to ensemble stellar processes.\ Because of this variability, one avenue for studying the effects of galactic outflows driven by AGN is to search for ``fossil'' AGN activity in addition to current activity \citep{2022MNRAS.516.4963I}.\ 
Seeking evidence of prior outflow events is thus vital for understanding the co-evolution of galaxies and their supermassive black holes (SMBHs) across cosmic time.  

At X-ray energies, Iron K$\alpha$ emission features are ubiquitous in AGN spectra.\ 
Sensitive studies with Chandra have now detected hard X-rays, including Fe~I~K$\alpha$, reaching out to kpc scales from the nucleus in nearby Compton-thick AGN \citep{2020ApJ...891..133J, 2018ApJ...855..131F, 2015ApJ...812..116B, 2020ApJ...900..164M}.\ 
This kpc-scale Fe I K$\alpha$ emission does not conform to the expected tens of pc for standard torus models \citep[e.g.,][]{1992ApJ...401...99P} and such scales are much larger than the observed sizes of tori, which ALMA and IR reverberation studies have shown to be $\sim0.10$ pc \citep[e.g.][]{2016ApJ...829L...7G, 2016ApJ...822L..10I, 2016ApJ...823L..12G, 2021ApJ...912..126L}.\ 
Spatially mapping neutral iron fluorescence emission at such distances from the SMBH challenges the long-held belief that these characteristic Fe K$\alpha$ lines originate purely via the excitation of nuclear or circumnuclear material from the AGN on small scales.\ 
The alignment of these large-scale features in many cases with jets and the optical narrow line regions suggests a direct connection with larger-scale AGN feedback.

The very nearby galaxy NGC 4945 (D$\sim$4 Mpc) offers an excellent opportunity to search for large-scale feedback.\ NGC 4945 is a starburst galaxy \citep{1990ApJS...74..833H, 2001A&A...372..463O} with a strong and variable, buried AGN \citep{2014ApJ...793...26P}.\ The galaxy is viewed at an almost edge-on inclination of $i > 80^{\circ}$.\ The galaxy disk contains significant dust lanes that obscure much of the activity in directions southeast of the nucleus, while the northwest side of the nucleus is relatively free of extinction \citep{2000A&A...357...24M, 10.3389/fspas.2017.00046}.\ The {\it Chandra} ACIS-S hard X-ray image of NGC 4945 shows a resolved, flattened, and clumpy region of cold reflection and neutral Fe K$\alpha$ on scales out to 230 pc \citep{2003ApJ...588..763D, 2012MNRAS.423L...6M}.  
Based on a distinct lack of UV ionizing photons, \citet{2003ApJ...588..763D} argue that the AGN must be fully covered in all directions by $N_{\rm{H}}\ge10^{22}-10^{23}$ cm$^{-2}$.\

NGC 4945 has a black hole mass of $1\times10^6$ M$_{\odot}$ \citep{1997ApJ...481L..23G}, which places it at the very low end of masses for Type 2 AGN \citep{2004ApJ...613..109H}.\ Being in this low mass range, we may not expect to find much at all in the way of AGN-driven feedback.\ In fact, while the circumnuclear region does have an $\sim$ arcminute-sized soft X-ray ``plume'', that emission has frequently been linked to a starburst-driven wind from the ongoing nuclear starburst \citep{2002MNRAS.335..241S} although contributions from AGN activity are not ruled out.

We discuss here the discovery with XMM-Newton of an extended Fe I K$\alpha$ nebulosity on a $\sim5$ to $\sim10$ kpc ($\sim 250'' - 480''$) scale in NGC 4945.\ 
XMM's large X-ray collecting area makes it sensitive to this extended emission beyond the circumnuclear size scales of 15$''$ to 20$''$ where Chandra is most sensitive.\ The heavy nuclear absorption (log[$N_{\rm{H}} / \rm{cm}^{-2} ] > 23$) blocks the central AGN fully below $\sim10$ keV, and enough to create a perfect scenario to allow a clear study of the extended emission.\ 

Throughout this work, we adopt H$_0$ = 67.8 km sec$^{-1}$ Mpc$^{-1}$ and z = 0.0019.\ The radial velocity of NGC 4945 within the local group is v = 300 km s$^{-1}$ \citep{1996AJ....111..794K}, which provides a luminosity distance of 4.40 Mpc and scale of 21 pc arcsec$^{-1}$ = 0.021 kpc arcsec $^{-1}$.\ We take the position of the H$_{2}$O megamaser as the location of the AGN \citep{1997ApJ...481L..23G}, at RA(2000) = 13h 05m 27.28s ; Dec(2000) = $-49^{\circ}$ $27'$ 58.0$''$
\footnote{We note that corrected to the reference frame defined by the 3K cosmic microwave background radiation, z = 0.0027, resulting in a luminosity distance of 11.9 Mpc and a cosmology-corrected scale of 58 pc arcsec$^{-1}$ = 3.45 kpc arcmin$^{-1}$.}

We present the observations and data processing steps in Section~\ref{sec:obs}, and the analysis and results from the X-ray imaging for Chandra, the EPIC pn and MOS1 in Section~\ref{sec:imaging}. In Section~\ref{sec:spectralanalysis}, we discuss the spatially-resolved X-ray spectral analysis for the EPIC pn.\ We discuss our results in the context of previous multiwavelength works that examined NGC 4945 in Section~\ref{sec:discussion}, with an examination of the images in \ref{subsec:image_disc}, a discussion of starburst properties in \ref{subsec:starburst}, and the origins of the Fe I K$\alpha$ line in \ref{subsec:fek_disc}.\ We provide our conclusions in Section~\ref{sec:conclusions}. 

\section{Observations and Data Reduction\label{sec:obs}} 
\subsection{EPIC pn}
To examine X-ray emission in the hard band, we focus here on the pn which is the best EPIC instrument for such a measurement due to its greater effective area than the MOS cameras.\ 
XMM was pointed at NGC 4945 in 2022 on July 5 (Obs.\ IDs 0903540101, 0903540301) and August 16 (Obs.\ IDs 0903540201, 0903540401).\ 
The time on target is split between 0903540101 and 0903540201 for a total duration of 170.9 ks.\ 
Unfortunately, 0903540201 is mostly flared out with only a few thousand seconds of good data, and so we focus here on the longest single observation, 0903540101, with an on-source duration of 124.9 ks.\

\begin{figure*}[ht!]
\plotone{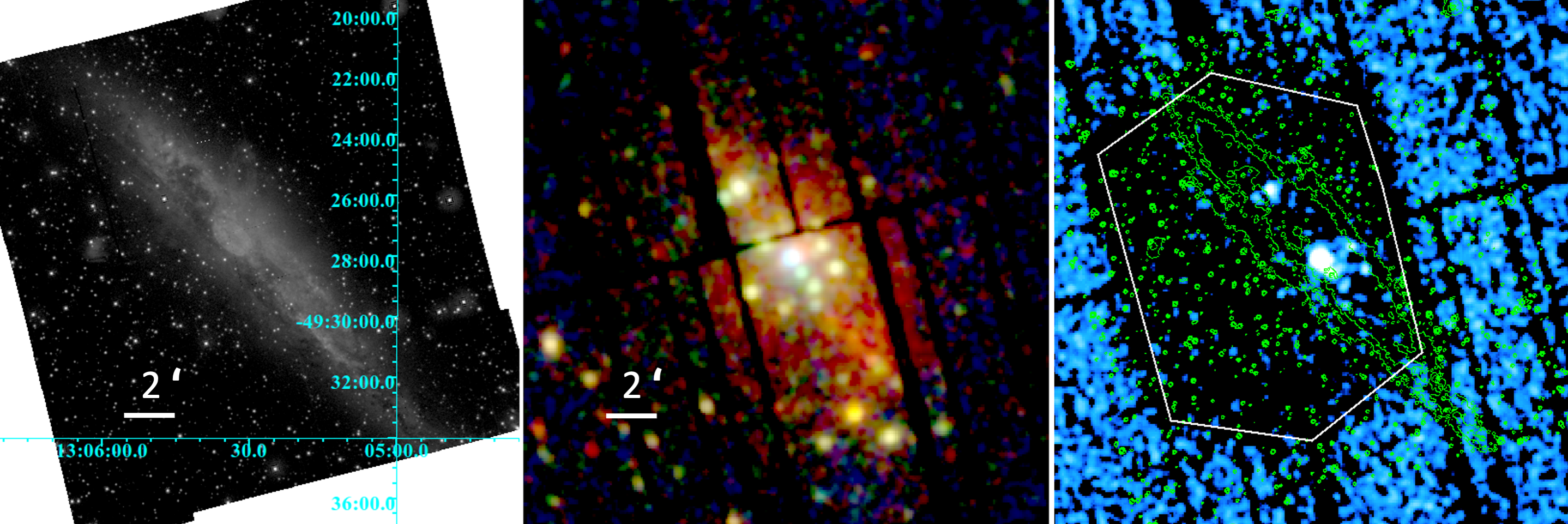}
\caption{\footnotesize{{\bf Left:} The XMM OM image in the B filter (400 to 500 nm).\ 
Artifacts are not corrected for but do not affect our analysis.\ {\bf Center:} A three-color image from the EPIC pn with red = 0.4 to 1.5 keV, green = 1.5 to 5.0 keV, and blue = 5 to 8 keV. The image has $4.1''\times4.1''$ pixels and is smoothed with a Gaussian with radius = $12''$.\ {\bf Right:} EPIC pn 7.2 to 8.0 keV image stretched to faint levels to show the central detector area with the lowest background (the ``Cu-hole'') in the pn (marked with the white hexagon).\ The green contours are the OM image overlaid.\ In all images, North is up and East is to the left.
}
\label{fig:fig1}}
\end{figure*} 

The pn data were examined to first order using the Science Analysis Software (SAS) version 20, and the data were reprocessed using the corresponding set of Current Calibration Files (CCFs).\ 
The \textsc{epchain} pipeline was used to produce event files with the most recent calibrations.\ 
These were then filtered using \textsc{evselect} with PATTERN $\le$ 4 (i.e., single- and double-pixel events) and FLAG == 0.\ to provide the most conservative screening criteria for spectral analysis. 

The X-ray instruments are susceptible to flaring, the result of soft protons in the magnetosphere being directed onto the detector by the mirrors, and this produces strong changes in the background.\ 
To check for these flares in the pn, we examined the light curve from the full field extracted with PATTERN == 0 and PI $\ge$ 10000 ($\ge10$ keV photons).\
This cut provided a mean count rate of $\sim0.19$ counts s$^{-1}$ and time intervals based on this energy band with rates less than 0.3 counts s$^{-1}$ were selected to filter out flares.\ The rate selection left 95.6 ks of good time.\ The events file was then filtered using \textsc{evselect} with various energy band selections to make cleaned images binned at the native pixel size of $4.1''$.\ We note for comparison with the MOS full-frame count rate quoted below, the pn full-frame count rate measured between 0.3 and 12 keV is $\sim4.4$ counts s$^{-1}$.

For spectral fitting, the next step was to check for pile-up.\ Pile-up occurs when more than one photon hits a pixel, or an adjacent one, before it is read out.\ 
Events are therefore summed within a readout cycle, leading to an artificially harder spectrum and lower count rate.\ 
To check for pile-up, data were extracted from a central region around the galaxy nucleus of radius $40''$ using the task \textsc{evselect} and  examined with task \textsc{epatplot}.\ Some pile-up is present at the level of a few percent.\ 
One method to correct for this is to filter out the multi-pixel events.\ 
However, since we wish to keep as many hard X-ray photons as possible, we used the standard method to create a response matrix for the pn that corrects for pile up.\ 
This includes running the \textsc{epic} reduction meta-task \textsc{epproc} to create an intermediate calibrated event file, corrected for X-ray loading effects.\ 
We then produced the source and background spectral files from this corrected events file.\ 
The response files including the pile-up correction were then generated from the raw events file.\ 
Finally, the source spectra were grouped to have a minimum of 25 counts per bin for spectral fitting in \textsc{xspec}.

\subsection{EPIC MOS}
To confirm structure in the pn and off-nuclear X-ray point sources, and to also compare with Chandra (Section \ref{subsubsec:chandra_ps}), we used the MOS data to supplement our imaging analysis.\ The MOS has an in-orbit FWHM smaller than the pn, and so while all cameras have the same half-energy widths of $\sim16-17''$, the MOS is better for distinguishing the peaks of the point sources.\ Also, we restricted ourselves to MOS1 for this analysis because the triangular shape of the PSF in MOS2 changes the shape of the central source region just enough that adding the two cameras together distorts the cleanness of the MOS1 image for this work.

The MOS1 data were examined similarly to the pn using the SAS version 20 and were reprocessed using the corresponding set of CCFs.\ The \textsc{emchain} pipeline was used to produce event files. These were then filtered using \textsc{evselect} with PATTERN $\le$ 12 and FLAG == 0.\ To check for background flares, we used the full field and examined the light curve with PI==300 to 12000.\ The mean count rate with this selection is $\sim1.5$ counts/s.\ Time intervals with count rates less than 1.9 cts/s were chosen to filter out flares, leaving 98.8 ks of good time.\ The events files were filtered using \textsc{evselect} with energy band selections to make cleaned images binned at the native MOS pixel size of $1.1''$.

\subsection{Optical Monitor}\label{sec:om}

XMM Optical Monitor (OM) images were collected in the V, B, U, UVW1, UVM2, and UVW2 filters, with exposure times of $\approx$ 10 ks (V), 15 ks (U), and 20 ks (B and UV filters) concurrently with the EPIC data.\ 
Data were processed and mosaic images were created using the SAS \textsc{omichain} command, with default settings.\ 
After processing, the images were examined and found to have some scattered light artifacts, which is common in this instrument: scattered light from bright, off-axis stars can be reflected by the detector housing onto the detector, producing a diffuse central doughnut, and/or oval loops or streaks of higher background that can be radially offset from the center.\ 
One such artifact was found near the center of the galaxy in all exposures; another was found about 4.5' from the galaxy's center near $\alpha$=13$^{h}05^{m}58^{s}$ and $\delta$=-49$^{\circ}$27$^{\prime}$47$^{\prime\prime}$ (J2000.0).\ 
These regions were disregarded in scientific analysis.
\linebreak

\subsection{Chandra ACIS}
To search for specific contributions from point sources beyond the AGN, we used Chandra's excellent imaging resolution which allows us to locate off-nuclear hard X-ray point sources that could confuse pn images and spectra.\ 
For the best statistics, we used the longest Chandra ACIS-S non-grating observation of NGC 4945, taken on 2014.12.10 (Obs. ID=14984, 130 ks).\ 
Data were reprocessed with the latest version of \textsc{ciao} \citep{2006SPIE.6270E..1VF} and the latest CALDB, and images and spectra created with CIAO tools.\ The good time interval exposure after reprocessing was 128.76 ks.
For source detection we used the tool \textsc{wavdetect}.\ 
Source and background spectra were extracted manually with \textsc{dmextract}, and response and arf files were generated for cases where point sources had sufficient hard X-ray counts to examine their spectra.

\section{Data Analysis and Results}
\subsection{Imaging\label{sec:imaging}}
\subsubsection{Large scale emission}
A three color pn image and its companion OM image are shown in Figure \ref{fig:fig1}.\
Comparing the two shows that the X-ray emitting material extends beyond the optical disk of the galaxy and into the galaxy halo.\ There is also an overall misalignment of the X-ray nebula with the disk plane where it extends further to the SE side than the NW side.\ 
The strongest sources of pn detector background in the 3-10 keV band are fluorescent Cu Ka lines that are also potentially variable \citep{2002A&A...389...93L}.\ Fortunately, the region of interest for this analysis falls in the ``Cu-hole'', a region of the pn detector with no instrumental Cu line (Figure \ref{fig:fig1}).\footnote{The Cu hole appears at PI==7700 to 8400 and PI==8700 to 9100.\ In detector coordinates the region is defined as DETX, DETY IN CIRCLE(-2260, -1090, 7500) and (DETX $> -8060$) and (DETX $< 3540$).} 
The MOS cameras do not have strong fluorescent lines above 2 keV. Thus, the background in the region of interest is low and non-variable (after filtering out proton flares). 

\begin{figure}[h!]
\includegraphics[width=1.0\linewidth]{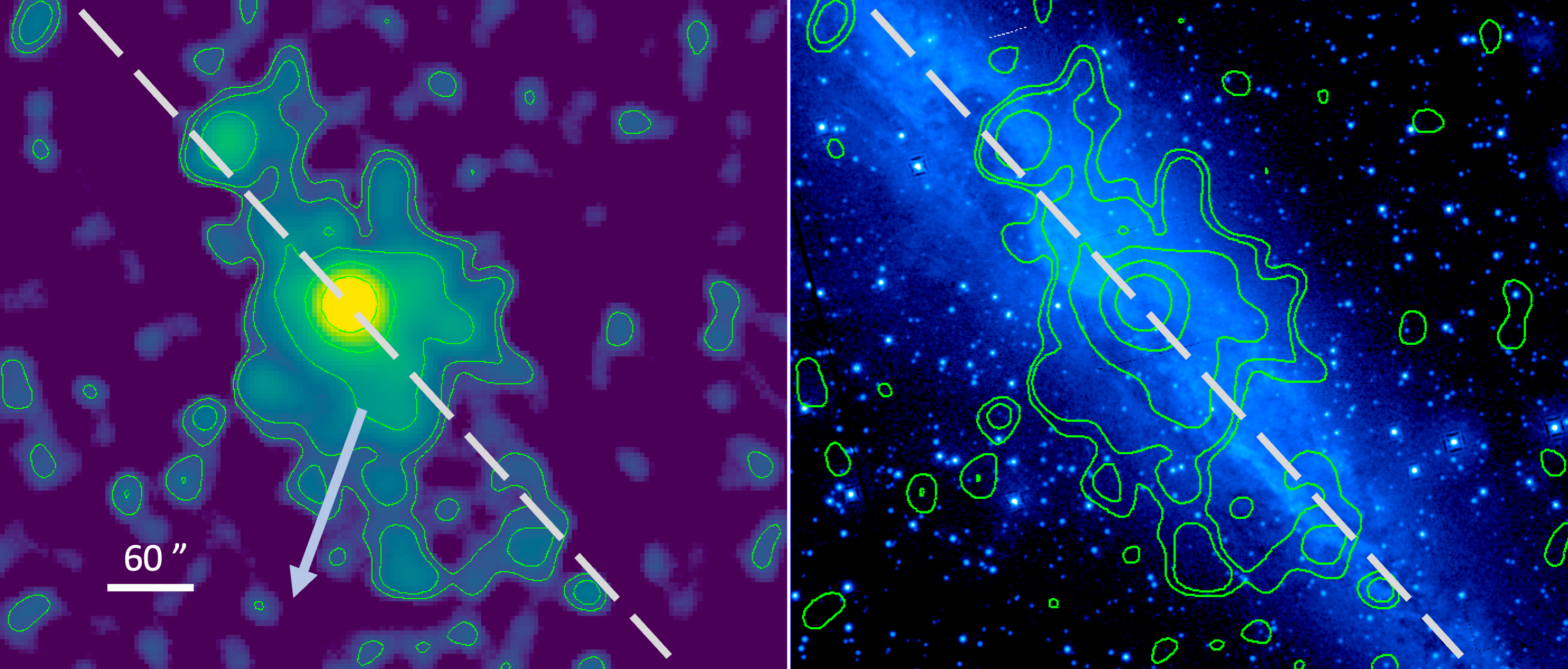}
\caption{\footnotesize{{\bf Left:} EPIC pn Fe K$\alpha$ plus continuum image (''Fe K'' image) and contours.\ The $4.1\times4.1''$ pixel image is smoothed with a Gaussian function with a radius of $25''$ at $3\sigma$.\ Contours are logarithmic starting at 0.13 counts with the lowest contour at $2.5\sigma$ above the mean background level. 
The dashed line indicates the optical location and position angle of the galaxy plane \citep{peterson1980}. The arrow indicates the general direction of the largest extension of soft X-ray emission out of the galaxy plane in projection (visible in Figure \ref{fig:fig1}).\ {\bf Right:} Overlay of the contours on the OM image in the B filter to show the relative orientation and size of the feature compared to the galaxy disk. 
}
\label{fig:fig2}}
\end{figure}

To examine the Fe I K$\alpha$ emission presumably associated with the AGN, we created an Fe K$\alpha$ + continuum image (``Fe K map'') within a narrowly defined 380 eV wide band from 6.2 to 6.58 keV.\ This map is shown separately and with contours overlaid on the OM B filter image in Figure \ref{fig:fig2}.\
The energy band is chosen to specifically avoid the $\sim$6.7 keV feature from collisionally ionized gas as much as possible (see Section \ref{sec:spectralanalysis}).\ 
We point out that this band captures the majority of the neutral or low-ionization Fe K$\alpha$ photons (Fe K I-XII) within the energy resolution of the pn, which is $\sim150$ eV at 6.4 keV.\ 

The first point of note is that the 6.4 keV emission region extends in projection $\sim480''$ (10.1 kpc)
along the plane of the galaxy, and to $\sim250''$ (5.3 kpc) above the plane.\
The second point is the overall extension to the NW and SE that begins to exit the galaxy plane in projection and in a direction angled 
with respect to the galaxy plane (direction of blue arrow in Figure~\ref{fig:fig2}).\ The extension to the SE especially seems to align with the overall direction of the extension to the SE for the soft X-ray emission seen in Figure~\ref{fig:fig1}.\ 
This spatial coincidence suggests that the soft X-rays and Fe K$\alpha$ emitting regions are (or have been) somehow connected.

\begin{figure}[h]
\includegraphics[width=1.0\linewidth]{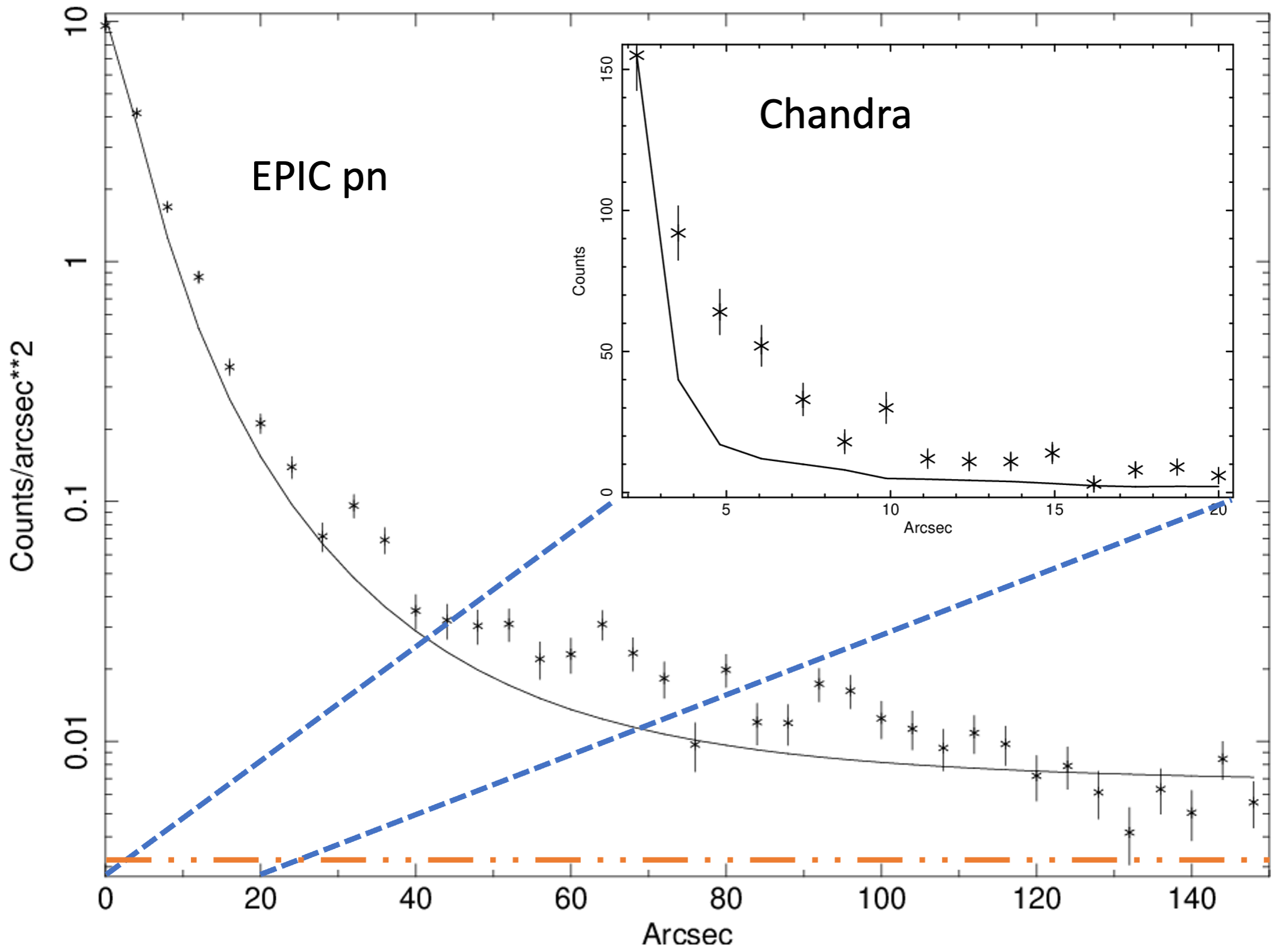}\\
\includegraphics[width=1.0\linewidth]{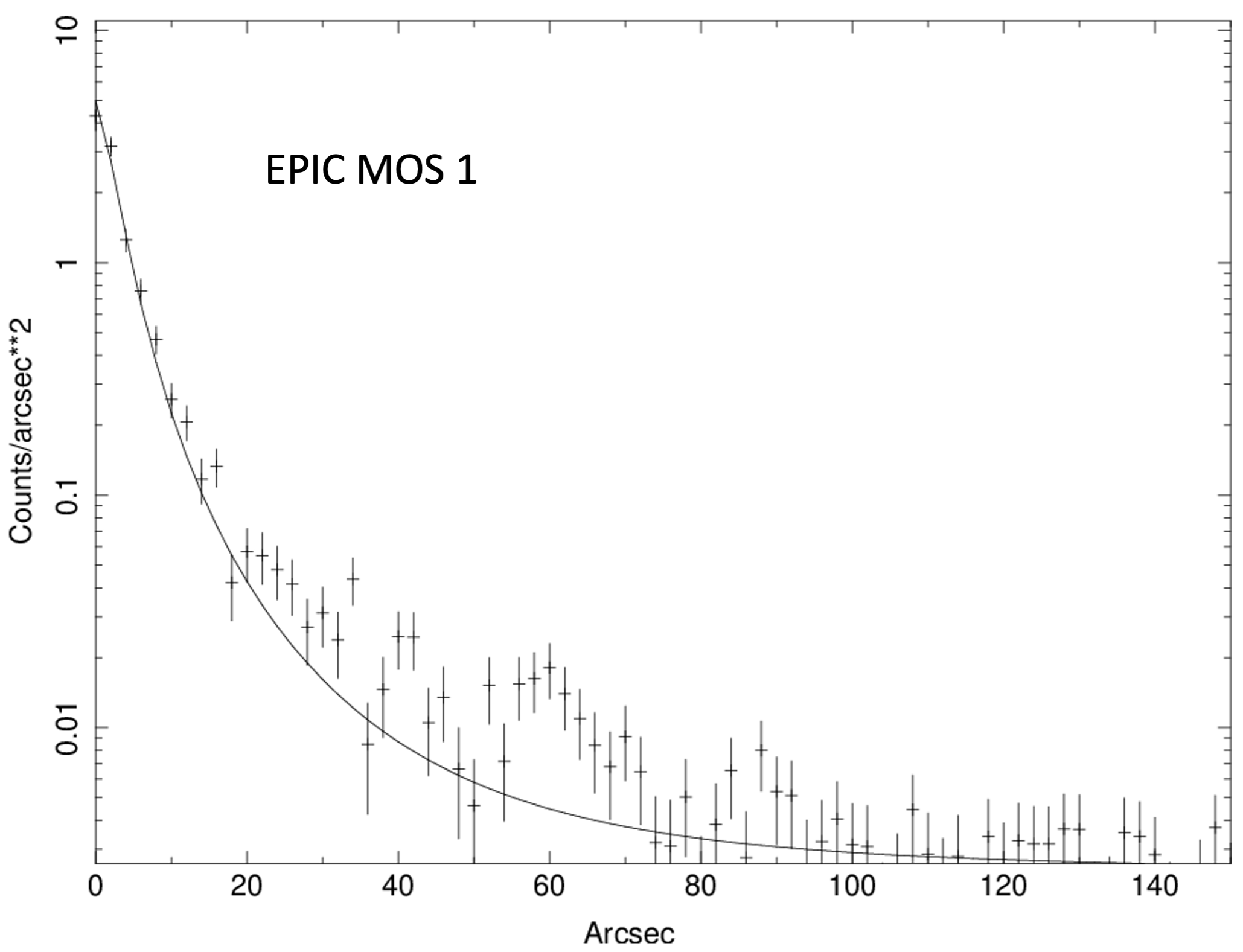}
\caption{\footnotesize{{\bf Top:} Radial profile in counts arcsec$^{-2}$ for the pn Fe K image compared to the EPIC pn PSF (solid black line).\ The orange dot-dashed line is the pn off-source background level.\ Inset is the radial profile for the Chandra Fe K image compared to the Chandra PSF.\ The y-axis is calculated differently and is in total counts (counts as a function of radial distance).\  {\bf Bottom:} Radial profile in counts arcsec$^{-2}$ for the MOS1 Fe K image compared to the MOS1 PSF (solid black line).}
\label{fig:fig3}}
\end{figure}

The strong AGN point source clearly dominates the pn image but also contributes to the circumnuclear emission.\ We suspect from Chandra observations of extended emission on a 10-$15''$ scale \citep{2012MNRAS.423L...6M} that the AGN will appear extended, even in the pn.\ For comparison, the XMM PSF has a $12.5''$ FWHM and a $16.6''$ half energy width at 5 keV.\footnote{XMM-Newton Users Handbook, Issue 2.21, 2023 (ESA: XMM-Newton SOC)}\ 
Adding complexity is the presence of possible extended hard X-ray continuum emission (Figure~\ref{fig:fig1}, right panel). Ideally we would subtract a continuum image to make an emission line map, but the non-uniformity of the pn detector background and effects of the chip gaps makes this difficult.\ Taking all of the caveats into account, we examined the data in several complimentary ways using both standard analysis tools and a direct comparison of counts.

\begin{figure*}[ht!]
\plotone{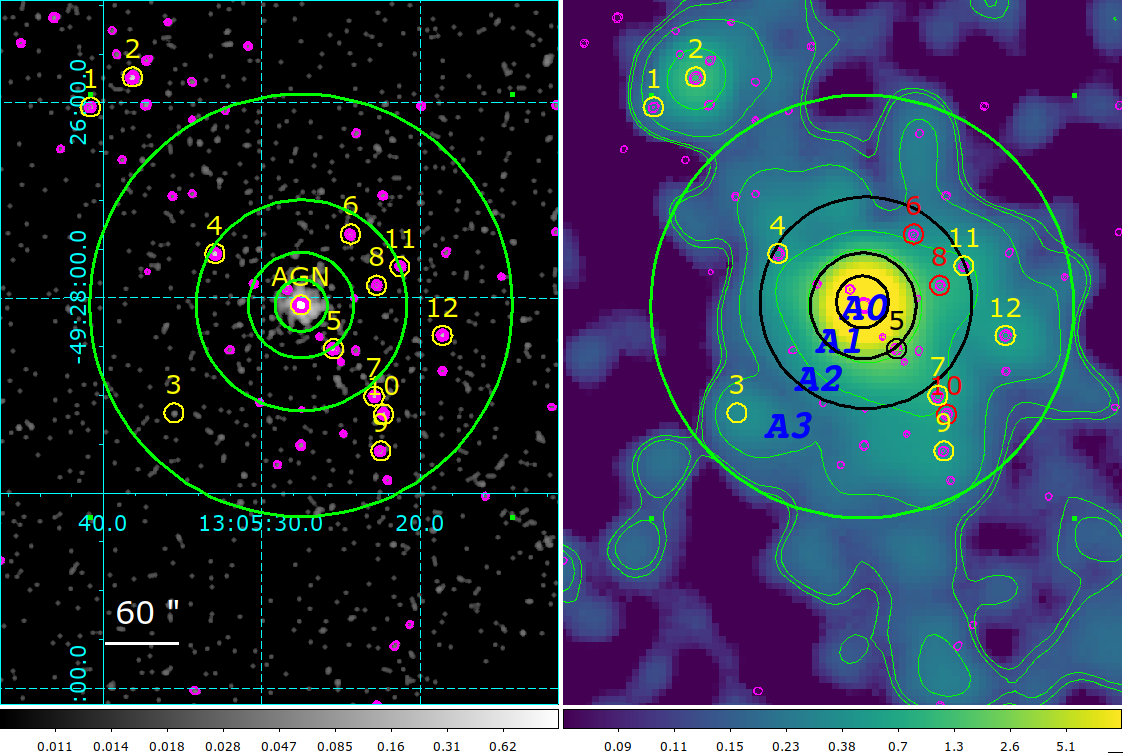}
\caption{\footnotesize{Chandra-detected X-ray point sources within the pn Fe K$\alpha$ nebulosity.\ {\bf Left:} The 2014 Chandra ACIS Fe K map is shown in greyscale with \textsc{wavdetect} contours overlayed (magenta).\ Image is smoothed in ds9 with a Gaussian function having a radius of $6''$.\ The large green circles that denote spatial and spectral extraction regions in this paper are drawn at radii of $16''$, $32.5''$, $65''$, and $130''$.\ These are the areas within which we sum up the total counts within regions -- discussed in the text -- that correspond to our pn spectral extraction regions (\ref{sec:spectralanalysis}).\ Yellow circles have a radius of $6''$ and define the twelve brightest Chandra sources (Table \ref{tab:pointsources}).\ The exception is source \#3 which is faint in Chandra but detected in the pn.\ {\bf Right:} The 2022 pn Fe K map with the Chandra contours and twelve brightest off-nuclear point source locations shown.\ Small circles have a radius of $6''$.\ Red circles indicate sources that have dimmed significantly or turned off between 2014 and 2022 (see Figure~\ref{fig:fig5}).\ The large annular regions are labeled in this panel as A0 to A3, and correspond to our Chandra point source summation regions (this section) and our pn spectral extraction regions (\ref{sec:spectralanalysis}).
}
\label{fig:fig4}}
\end{figure*}

\subsubsection{{Radial profiles}\label{subsubsec:radial}}

To determine how much of the pn emission is specifically from the point-like AGN core, we constructed a radial profile of the Fe K map with the SAS task \textsc{eradial} using a King profile at an energy of 6.4 keV (Figure~\ref{fig:fig3}).\ This plot shows significant counts in excess of the modeled PSF from about $16''$ to $120''$.\ We note that there is a few percentage of pile-up in the core of the image but this does not have a significant effect on matching the shape to the PSF at radii of greater than $4''$ and so we can ignore the effects of pile-up for this analysis.\ One possible issue with this technique is that \textsc{eradial} doesn't extend beyond $150''$ because that's how far the PSF model extends.\ The task fits a profile by scaling the PSF to the peak plus a constant background term, which in our case was fitted at 0.0031 counts arcsec$^{-2}$.\ In examining image directly, this level seems correct and is consistent with the off-source background in several regions outside the nebulosity.\ 

From the radial profile plot and comparing the brightness levels, we can estimate that almost $40\%$ of the AGN emission is extended even at $\sim16''$.\ This circumnuclear fraction is on the same order as that seen in the Chandra data \citep{2012MNRAS.423L...6M}.\ We next examined the extended emission in Chandra around the AGN core to confirm our suspicion that the pn is observing this circumnuclear extent.\ 
From the 6.4 keV image, we used the \textsc{ciao} task \textsc{dmextract} with annular regions defined in DS9 from $0.5''$ to $20''$ to extract a radial profile.\ 
We then created a basic PSF using the task \textsc{arfcorr}, which generates an image with a simulated circularly symmetric PSF, generated to have the correct encircled count fraction at each radius.\ 
We extracted a radial profile from this image at the same binning and compared the two (Figure~\ref{fig:fig3}, top, inset).\ 
The Chandra emission is seen to $\sim15''$.\ Within $16''$ we added together the total counts to obtain the extended fraction in the Chandra map.\ There are 512 total Fe K plus background counts and the model PSF predicts 272, which gives 240 in the extended nuclear component, or a fraction of $\sim47\%$ -- consistent with \citet{2012MNRAS.423L...6M} and roughly consistent with our pn estimate of a circumnuclear fraction of $\sim40\%$.

We then examined spatial scales from $16''$ to $130''$.\ We note that due to the chip gaps that are not corrected for, a radial profile is not the best way to approximate this extended region in the pn (see regions drawn in Figure \ref{fig:fig4}).\ Integrating the PSF model prediction across the region from $16''$ to $130''$ for the pn yields 0.94 counts arcsec$^{-2}$ while integrating the pn source emission yields 1.36 counts arcsec$^{-2}$.\ This suggests that roughly $\sim30\%$ of the source flux in total lies in excess of what we expect from the AGN between $16''$ to $130''$.

From the MOS1 radial profile (Figure \ref{fig:fig3}), if we focus only on the $60''$ radius, $\sim80\%$ of the emission is in excess of the AGN.\ From the pn at $60''$ about $50\%$ of the emission is in excess of the AGN.\ The exact fraction for a direct comparison between the MOS and pn at any radius would depend on proper exposure corrections, and also modeling out the point sources (see Section \ref{subsubsec:psf_subtract}).\ We conclude from the radial profiles that a significant fraction of the extended Fe K band flux is in excess of that expected from the bright AGN point source.\ We relate this to the Chandra results on diffuse emission vs.\ point sources in the following section. 

\begin{figure*}[h]
\includegraphics[width=0.95\linewidth]{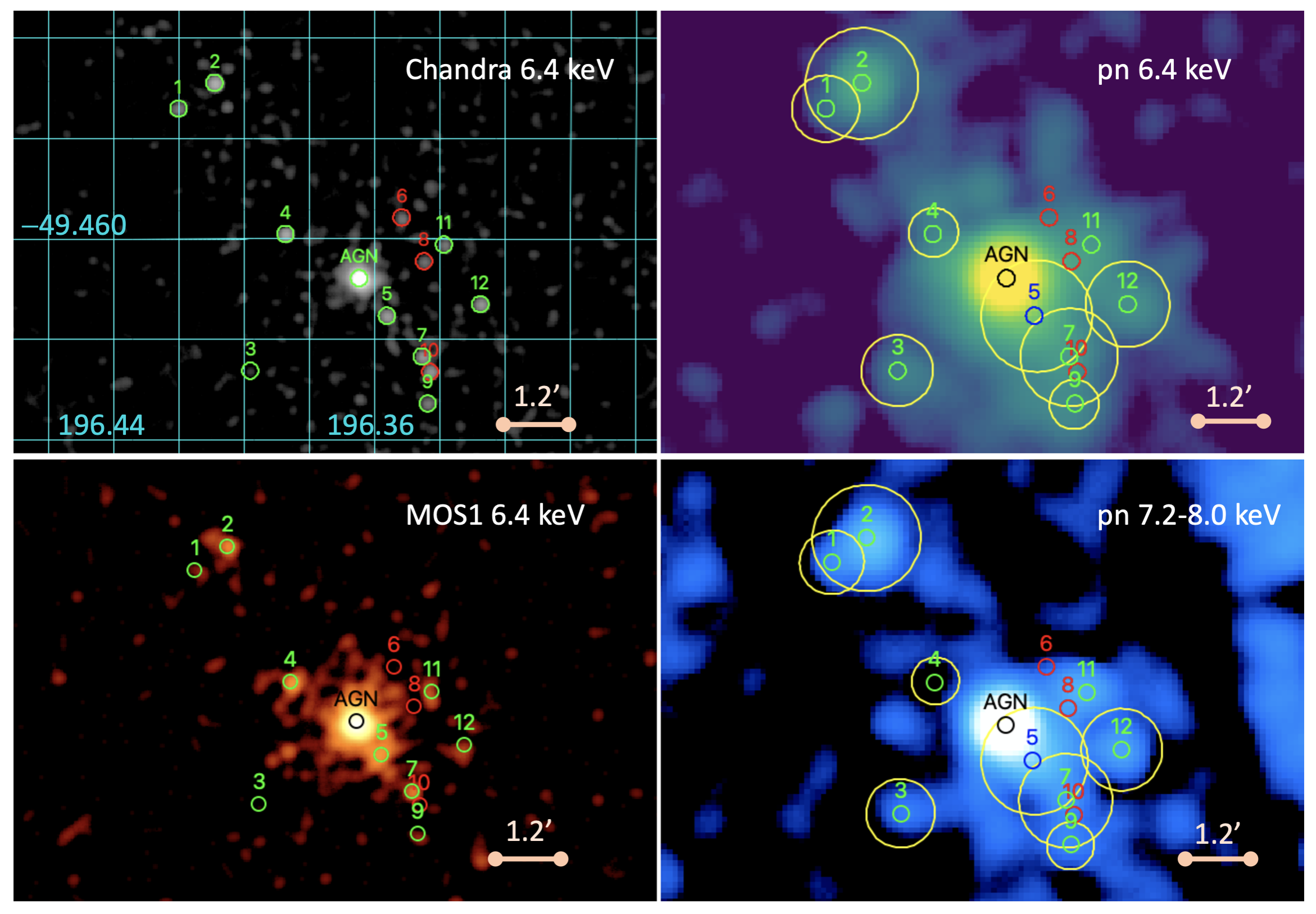}
\caption{\footnotesize{Comparison of narrow-band images with Chandra-detected and pn-detected source regions.\ {\bf Top Left:} Chandra Fe K map.\ Image is smoothed in ds9 with a Gaussian function having a radius of $6''$.\  Point sources found using \textsc{wavdetect} with a significance of $\ge5 \sigma$ (with the exception of \#3) are denoted with green and red circles.\ Red circles indicate those not clearly seen in the MOS (panel 3) and assumed to fall below the local background in the pn image.\ {\bf Top Right;} EPIC pn Fe K map smoothed with a Gaussian of $25''$ radius.\ Chandra regions are overlayed.\ The yellow ellipses denote the largest 98$\%$ encircled energy regions for the pn-detected point sources found using \textsc{edetectchain} on the hard band image (7-10 keV) and/or the 6.4 keV band image.\ {\bf Bottom Left:} EPIC MOS1 Fe K map smoothed with a Gaussian of $6''$ radius with the Chandra regions overlayed.\ Three sources are not detected.\ {\bf Bottom Right} pn 7.2 to 8.0 keV map smoothed with a Gaussian of $25''$ radius with regions overlaid to show the coincidence of point sources with this continuum band map.\ A version of this panel is shown in Figure~\ref{fig:fig1} with less smoothing.\ Source \#4 falls into the chip gap but is seen in the MOS (panel 3).
}
\label{fig:fig5}}
\end{figure*}

\subsubsection{Handling point sources from the Chandra image\label{subsubsec:chandra_ps}}

Moving beyond the central AGN-dominated region to larger scales we examine the contribution to extended emission from point sources.\ Key in measuring the full extent and fraction of Fe K emission that comprises the nebulosity is properly accounting for the many known X-ray point sources within the galaxy \citep{2016ApJS..224...40W,2004ApJ...602..231C,2012ApJ...756...27L}.\ 
The Chandra 6.4 keV plus continuum image is shown in Figure~\ref{fig:fig4}.\ 
To measure the Chandra point sources within the pn nebulosity (Figure~\ref{fig:fig2}) we ran the \textsc{ciao} task \textsc{wavdetect} on this image.\ Plotted in Figure~\ref{fig:fig4} is the Chandra Fe K map smoothed with a Gaussian of radius = $6''$ with the \textsc{wavdetect} contours overlaid for sources detected above a level of $5\sigma$ significance above the local diffuse background.\ 
We define two categories of point sources in this map, which we classify for our purposes as bright and faint: bright sources have 6 or more counts in 128.76 ks; faint sources generally have 2 to 3 counts.\ We list the ''bright'' sources in Table \ref{tab:pointsources}.\ 
All of these sources appear in the catalog of \citet{2016ApJS..224...40W}, with many listed in \citet{2004ApJ...602..231C} and also the XMM catalog of \citet{2012ApJ...756...27L}.\ The outlier is source \#3, which is faint but we include because it may be detected with the pn (see below).

Because we are dealing with photon counting statistics in such a narrow energy band it is important to have an estimate of the background level and count rate uncertainty.\ Clearly from Figure~\ref{fig:fig1}, there is a strong local X-ray background seen with XMM in the galaxy, but it is difficult to measure very faint diffuse emission with Chandra.\ Examining the wavdetect output for the regions around the AGN but outside the circumnuclear region (within area A1 and A2 in Figure~\ref{fig:fig4}) we determine a mean local background level within a $6''$ radius of 1 to $2 \pm1$ counts and an off-galaxy level of $\sim0.5 \pm0.5$ counts.\ This provides a comparison for determining whether a source is clearly detected above the local diffuse (galaxy) background, as well as an estimate of the average brightness of the diffuse galaxy emission compared to the sky plus detector background.

We next examine each point source individually in the Chandra image using DS9 and the wavdetect output to determine the relation of source plus background count rate to the AGN plus background count rate.\ We split the image into large annular regions to add together the source plus background counts within individual point source regions ($6''$ radii - Table \ref{tab:pointsources}) and within the full annuli, to determine what fraction of those total counts can be ascribed to individual point sources.\ The annular regions discussed here are shown in Figure~\ref{fig:fig4} and have radii of $0 - 16''$ (annulus 0 - A0), $16'' - 32.5''$ (annulus 1 - A1), $32.5'' - 65''$ (annulus 2 - A2), and $65'' - 130''$ (annulus 3 - A3).

\begin{deluxetable*}{llllcccccc}
\tabletypesize{\scriptsize}
\tablewidth{0pt} 
\tablecaption{Off-nuclear X-ray Point Sources in the Chandra and MOS1 Fe K$\alpha$ Maps\label{tab:pointsources}}
\tablehead{
\colhead{Number} & \colhead{RA (2000)}& \colhead{Dec. (2000)} & \colhead{Name} & \colhead{Annulus} &
\colhead{Chandra Counts} & \colhead{Chandra SB$^a$} & \colhead{$\%$ of AGN}& \colhead{MOS1 SB$^a$} & \colhead{$\%$ of AGN} \\
\colhead{} & \colhead{}& \colhead{} & \colhead{} & &
\colhead{ds9 / wavdetect} & \colhead{(counts arcsec$^{-2}$)} &  & \colhead{(counts arcsec$^{-2}$)} &  \\
} 
\colnumbers
\startdata
AGN & 13 05 27.3 & $-$49 27 58.0  &  NGC 4945            & A0 & $658\pm25$ / $455\pm24$  & 6.43  & 100   & 1.53 & 100 \\
1$^b$ & 13 05 40.8  & $-$49 26 02.7   & CXOGSG J130540.7-492603  & -- & $9^{+4}_{-3}$ / $6.3^{+3.7}_{-2.5}$ & 0.08   & $\approx1.2$  & 0.012 & 1.0  \\ 
2 & 13 05 38.1  & $-$49 25 45.6  & CXOGSG-J130538.0-492545 & -- & $32\pm6$ / $24.8\pm5.3$ & 0.32   & $\approx5.0$  & 0.056 & 3.7  \\
3 & 13 05 35.5 & $-$49 29 10.8  & CXOGSG-J130535.4-492911 & A3 & $2^{+3}_{-1}$ / $2^{+2.7}_{-1.3}$ & 0.02 & $\approx0.3$  & 0.006 & 0.4   \\
4$^c$ & 13 05 32.9 & $-$49 27 33.6  & CXOGSG-J130532.8-492733 & A2 & $25\pm5$ / $23\pm5$ & 0.23  & $\approx3.6$  & 0.072 & 4.7 \\
5 & 13 05 25.4 & $-$49 28 31.6  & CXOGSG-J130525.4-492832 & A1/A2 & $18^{+5}_{-4}$ / $16^{+5.1}_{-4.0}$ & 0.17  & $\approx2.6$  & 0.083 & 5.4  \\
6 & 13 05 24.3 & $-$49 27 20.8  & CXOGSG-J130524.3-492721 & A2 & $7^{+4}_{-3}$ / $6^{+3.6}_{-2.4}$ & 0.07  & $\approx1.0$  & {\underline {0.006}} & {\underline {0.4}}  \\
7$^d$ & 13 05 22.8 & $-$49 29 01.0  & CXOGSG J130522.8-492901 & A3 & $16^{+5}_{-4}$ / $14.5^{+4.9}_{-3.8}$ & 0.14 & $\approx2.1$  & 0.045 & 2.9   \\
8 & 13 05 22.7 & $-$49 27 52.5  & CXOGSG J130522.6-492752  & A2 & $6^{+4}_{-2}$ / $4.8^{+3.4}_{-2.1}$ & 0.06  & $\approx1.0$  & {\underline {0.003}} & {\underline {0.2}}  \\
9 & 13 05 22.5 & $-$49 29 33.7  & CXOGSG-J130522.5-492934 & A3 & $8^{+4}_{-3}$ / $5.7^{+3.5}_{-2.3}$ & 0.07 & $\approx1.1$  & 0.015 & 1.0  \\
10 & 13 05 22.3& $-$49 29 12.0  &  CXOGSG J130522.2-492912  & A3 & $16^{+5}_{-4}$ / $13.4^{+4.8}_{-3.6}$ & 0.15 & $\approx2.3$ & {\underline {0.005}} & {\underline {0.3}}   \\
11 & 13 05 21.1& $-$49 27 40.5 & CXOGSG J130521.1-492740 & A2/A3 & $8^{+4}_{-3}$ / $6.6^{+3.7}_{-2.5}$ & 0.07 & $\approx1.1$ & 0.028 & 1.8  \\ 
12 & 13 05 18.5  & $-$49 28 23.0   & CXOGSG-J130518.5-492823 & A3 & $15^{+5}_{-4}$ / $13.3^{+4.7}_{-3.6}$ & 0.15 & $\approx2.0$  & 0.021 & 1.4  \\ 
\enddata
\tablecomments{Chandra names are from \citet{wang2016}.\ $^a$ SB is surface brightness within a $6''$ radius.\ $^b$Source 1 is also 2XMM J130540.7-492602 \citep{2012ApJ...756...27L}.\ $^c$Source 4 is a ULX \citep{swartz2004}.\ $^d$Source 7 is NGC 4945 X-1, a well known, variable ULX \citep{1996MNRAS.281L..41B}. We quote Gaussian error bounds for sources with $>20$ counts and Poisson error bounds \citep{1986ApJ...303..336G} for sources with $<20$ counts.\ Underlined numbers indicate point sources with significant variability between 2014 and 2022.}
\end{deluxetable*}

\vspace{-0.5cm}

In the following we assume photon statistics counting errors, but we do not propagate the errors because we are using the Chandra data as a consistency check rather than a detailed analysis of these sources.\ The relative error bars for individual sources are listed in Table \ref{tab:pointsources}.

Starting with A1, there are $\sim72$ total counts: one bright source has $\sim10$ counts and two faint sources add up to $\sim5$ counts.\ This equates to $\sim21\%$ of the counts from point sources.\ A2 has $\sim176$ total counts; five bright sources provide $\sim50$ counts and three faint sources provide $\sim10$ counts.\ Therefore, $\sim34\%$ of the counts in A2 are from point sources.\ A3 has $\sim455$ total counts: $\sim52$ counts are from six full or partial bright sources while $\sim36$ counts arise from twelve faint sources.\ Thus $\sim19\%$ of the counts in A3 are from point sources.\ Averaging out, about $25\%$ of the counts in the Chandra field between $16''$ and $130''$ are from point sources, with over half of those coming from the nine brightest X-ray point sources. 

Looking at it another way, $\sim75\%$ of the Fe K map photons between $16''$ and $130''$ from nucleus in the Chandra image are diffuse counts.\ This diffuse emission would be a combination of background, galaxy continuum emission, line emission, and very faint point sources.\ From our estimate of the background levels in the image, about 35\% of these counts are due to sky plus detector background, which leaves a $\sim40\%$ extended fraction in Chandra on large scales.

Our overall estimate from the pn, MOS1 and Chandra is that $\sim30\%$ to $\sim40\%$ of the Fe K + continuum lies in excess of the AGN outside of the immediate circumnuclear region between radii of $16''$ to $130''$.\ It is difficult to tease out the fraction of extended emission that would be purely line emission and also the Chandra result was obtained by {\it excluding} the bright point sources while the point sources are still included for the EPIC radial profiles.\ To provide more clarity we produce a point source subtracted pn image. 

\subsubsection{Handling point sources from the pn image}

For the pn, we searched for point sources with the task \textsc{edetectchain}.\ There are many point sources in and around the galaxy, but we restrict this work to the area with EPIC X-ray sources within the Fe K nebulosity plus those either confirmed with or detected separately with Chandra (Figure~\ref{fig:fig4}, Table \ref{tab:pointsources}).\ The pn results are shown in Figure~\ref{fig:fig5}.\ Yellow circles are the $98\%$ encircled energy regions for pn point sources within the nebulosity found using \textsc{edetectchain} on either the pn 7 to 10 keV image or the pn-Fe K map.\  There are eight point sources in the pn in this region.\ We overlay these also on the 7.2 to 8 keV image.\ Chandra source number 4 lies within the chip gap and is not detected.\ The brightest point sources that we consider below for PSF subtraction are numbered in Figure~\ref{fig:fig4} and Figure~\ref{fig:fig5}.

\subsubsection{Point source variability}

We now discuss the effects of variability between 2014 and 2022.\ The AGN in NGC 4945 is historically variable at energies above 10 keV \citep{2014ApJ...793...26P, 2004AIPC..714..190M}.\ But below 10 keV, the AGN is dominated by the reflection component.\ In this case the 2 to 10 keV fluxes for the central AGN are about the same between 2014 and 2022 at $\sim1.9\times10^{-12}$ ergs cm$^{-2}$ s$^{-1}$. Comparing the brightness of the non-nuclear point sources as a ratio to the AGN brightness allows solving for relative strengths of the point sources in two distinct data sets (XMM and Chandra).\ Next we can obtain a scaling factor for PSF subtraction.\ Thus we rely on the much better Chandra imaging capability to scale source peak fluxes in confused regions for the lower-resolution and higher background XMM data.

Any variability of bright point sources will have a significant effect on our estimate of their fractional contribution when comparing Chandra with XMM.\ To account for variability and starting with Chandra, we estimate the contribution of each point source relative to the AGN core in Chandra {\it at that time}.\
We obtained the number of counts, including statistical errors, using both ds9 and wavdetect.\ The surface brightness measurements for the AGN and all of the bright point sources within a region of $6''$ are listed in Table \ref{tab:pointsources}.\
Based on the Chandra PSF, this encircled energy contains $\ge95\%$ of the source (Figure~\ref{fig:fig3}).\ We then divided the surface brightness for each source by that measured for the AGN to obtain an approximate scaling factor of the relative brightness (Table \ref{tab:pointsources}, column 7).\ This scaling factor is used for our PSF subtraction estimates below. We note that from this method there are typical variations by up to a factor of two.

\subsubsection{Adding the MOS for cross-check}
We returned to the EPIC images to check for consistency between Chandra and the MOS1,  the latter of which provides a FWHM resolution slightly better than the pn due to a smaller pixel size. The ability of the MOS to pick out the point sources better than the pn, and the fact that this is a simultaneous measurement, allows an additional step in the EPIC flux comparisons for the Chandra point sources.

To estimate how bright each Chandra point source is relative to the AGN in the MOS, we used the Fe K line map for the MOS1 (Figure~\ref{fig:fig5}).\ To be consistent, we again obtained the surface brightness within a $6''$ radius for the AGN and all of the Chandra point source locations (Table \ref{tab:pointsources}).\ We divided the surface brightness for each source region by that measured for the AGN and obtained the MOS scaling factor (Table \ref{tab:pointsources}, column 9).\ We note that the derived scaling factors for the point sources with respect to the AGN point source are comparable for the Chandra and MOS Fe K images.\ This gives us confidence in this simple technique even with the statistical uncertainties for small number photon counting statistics.\ The exceptions within this method are the three sources (sources 6, 8 and 10 in Table \ref{tab:pointsources}) that are not obviously detected in the MOS exposure, being at fluxes less than the local background level.\ In our original Chandra point source flux estimates, these provided about half of the counts in region A2.\ On the other hand, source \#4, which fell into the chip gap in the pn is clearly detected in the MOS1 image. 

We make special note here that we have examined the 2014 Chandra image in detail and do not detect any significant Fe K map X-ray point sources within our hook region (see Section \ref{subsubsec:hook} and Figure \ref{fig:fig7} below).\ There is a soft X-ray ROSAT PSPC point source at 13h 05m 10.3s, $-49$ $31'$ $19''$ \citep{2000A&A...356..463G} that falls within the SW tail of the hook, but we do not see a clear corresponding point source in the MOS1 or the pn Fe K maps.\ This is not to say that there could not have been a significant continuum source in that region in the past.
 
Our results indicate that for the pn, we must  account for point sources 1, 2, 3, 4, 5, 7, 9, 11, and 12 and we include these sources in our PSF-subtracted image analysis in the next section.

\begin{figure*}[h]
\includegraphics[width=0.99\linewidth]{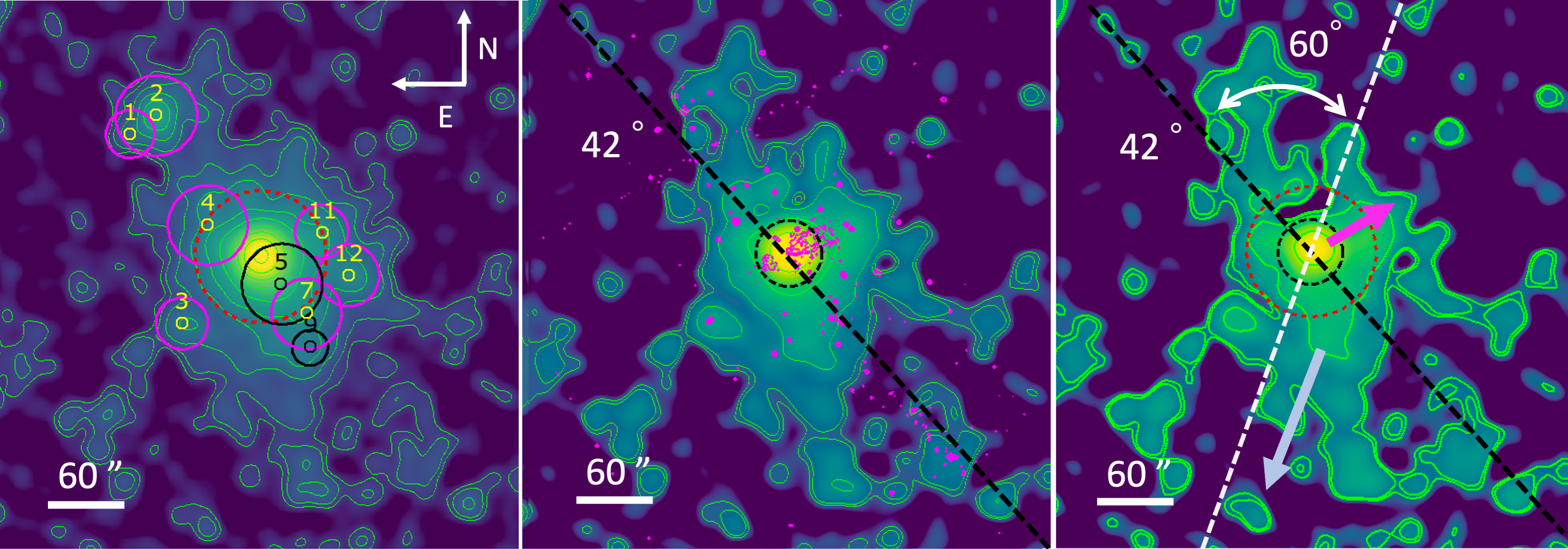}
\caption{\footnotesize{{\bf Left:} The pn Fe K image (Figure \ref{fig:fig2}) exposure-corrected to compare with the modeled PSF.\ The original image was re-binned with $1\times1''$ pixels, corrected, and then blocked by a factor of 4 in ds9 to display $4\times4''$ pixels.\ The image is smoothed with a Gaussian with a radius of $20''$ at $10\sigma$.\ Contours are in a log scale starting at $2\sigma$ above the background level with ten contours from 0.0075 to 1.63 counts.\ The large dashed circle is the $98\%$ encircled energy region for the central AGN point source.\ Other large circles are the XMM pn regions for X-ray sources that are being modeled; small circles are the Chandra regions.\ {\bf Center:} The PSF-subtracted Fe K map that still contains the circumnuclear diffuse Fe K emission; created using the technique explained in the text.\ This image is smoothed with a Gaussian with radius of $19''$ at $9.5\sigma$.\ Contours are in a log scale drawn starting at $2.5\sigma$ above the median background level with five contours from 0.0074 to 0.2 counts.\ The black dashed circle represents the boundary of region A1 (where the circumnuclear X-ray emission dominates - defined with a $32.5''$ radius - Figure \ref{fig:fig4}).\ The black dashed line is the position angle of the optical galaxy disk \citep[42$^{\circ}$, rotating from North to East;][]{ peterson1980}.\ The magenta contours are the 0.5 to 1.5 keV Chandra contours overlaid to illustrate the nuclear soft X-ray outflow \citep{2002MNRAS.335..241S}.\ The point sources here are soft X-ray sources and may have no relation to the hard X-ray sources discussed in this paper.\ {\bf Right:} The Fe K map that results from the additional removal of the circumnuclear extended component using a Gaussian blur on the autocorrelation function as described in the text.\ This image is smoothed with a Gaussian with radius of $20''$ at $10\sigma$.\ Four contours are drawn in a log scale starting at $2\sigma$ above the median background level from 0.0072 to 0.08.\ The white dashed line is our estimated rotation angle of $60^{\circ}$ for the PSF-subtracted Fe K nebulosity in projection with respect to the galaxy disk.\ The red dashed circle is the same as in the left-hand panel.\ The blue arrow is the same as in  Figure \ref{fig:fig2}.\ The magenta arrow indicates the approximate direction of the nuclear Chandra outflow in projection. }
}
\label{fig:fig6}
\end{figure*}

\subsubsection{PSF subtracted image}\label{subsubsec:psf_subtract}

Our goal is to obtain the best possible estimate of a PSF subtracted EPIC pn image of the Fe I K$\alpha$ nebulosity.\ By using the Chandra data to estimate how strong the off-nuclear sources are relative to the AGN and validating our estimates with the MOS image, we have a way to scale a modeled XMM PSF to each of these sources and subtract them from the original pn Fe K map.\ To prepare, we modeled the telescope PSF using the SAS task \textsc{psfgen}, at an energy of 6.4 keV centered on the nucleus ($\alpha$=13:05:41; $\delta$=$-$49:26:01) with level==EXTENDED.\ To match the PSF data, we also performed this work on a pn Fe K image with subpixel resolution of $1\times1''$ pixels.

The pn Fe K image was first corrected for CCD gaps by using the SAS task \textsc{eexpmap} to create an exposure map and then using this map to create a mask.\ This mask was then dilated by 1 pixel, and the CCD gaps were filled in within the FITs file using the \textsc{inpaint} algorithm built into Mathematica using the default texture matching settings.\ This effectively creates an exposure-corrected image and matches both the noise and background of the active image within the CCD.\ Then the PSF was cross-correlated with the result to produce a somewhat blurred image with minimal noise (Figure \ref{fig:fig6}, {\bf left panel}).\ This type of matched filter technique is sometimes used for astronomical images \citep{Huo_2015,Nir_2018} as a noise reduction technique and for point source detection.\ 

The XMM PSF was used as a base model for all the point sources.\ However, since the original image is naturally correlated with the PSF, the scaled model for the point sources is the autocorrelation function (which is the PSF correlated with the PSF).\ The autocorrelation function was first aligned to the AGN core by cross-correlating with the image described above and locating the absolute maximum.\ The other sources were located using the high-resolution Chandra data and matching the locations exactly in the XMM pn image. 

Having determined an approximate scaling factor for all of the point sources (Table \ref{tab:pointsources}), we were able to then scale the PSF and subtract the X-ray point sources according to their (source + background) scaling factor.\
The autocorrelation function was also subtracted from the AGN using a particular scale where the subtraction minimizes the variance about the peak to estimate the precise point source (Figure \ref{fig:fig6}, center panel).\ However, this subtraction still leaves residual AGN emission because this nuclear emission profile is broader than the XMM PSF (see again the PSF comparison in Figure \ref{fig:fig3}).\ This extended component was next estimated via this center panel image by using a Gaussian blur on the autocorrelation function with a radius of 5 pixels, which seemed to approximately match the shape of this residual extended emission in the pn.\ The final PSF-subtracted image is shown in Figure \ref{fig:fig6} (right panel). 

From the final PSF-subtracted image, and comparing with the original, it is still unclear how well the subtraction works in the center so we only consider areas outside of $32.5''$ from the core.\ The region we consider corresponds to our A2 and A3 plus the hook and adding the remaining extended emission.\ Taking a region that is a box 7.3 x 8.8 arcmin ($440'' \times 526''$) centered on the nucleus and excluding emission within a $65''$ radius, we find that approximately 32\% of the Fe K map (line plus continuum plus background) counts remain after point source subtraction.\ If we take the same box and exclude the area within $32.5''$, $\sim36\%$ is diffuse emission.\ These numbers are consistent with the rest of our analysis and so we feel confident that our image subtraction technique has not introduced significant uncertainties.\ 

We now see that the morphology of the nebulosity with the extension to the SE (see Figure \ref{fig:fig2}) and the offset orientation with respect to the galaxy disk are revealed with more clarity.\ For the optical major axis of NGC 4945, we adopt a position angle (PA) of 42$^{\circ}$ \citep{peterson1980}.\ The galaxy disk is represented in Figure \ref{fig:fig3}, panels 2 and 3, with the black dashed line.\ In the final image (panel 3) we have added the white dashed line to represent our $60^{\circ}$ rotation inferred for the nebulosity, which matches with our original visual estimate of the alignment in the direction of the greatest area of soft X-ray extended emission to the south (Figure \ref{fig:fig1}, Figure \ref{fig:fig2}).\ For a final best estimate of the fraction of the line emission that is extended, we used the 7.3 x 8.8 arcmin region to obtain the background counts in an off-galaxy region.\ We note that this likely represents a higher 6-7 keV background than the area in the Cu-hole, but subtracting this background level leaves a ``true" extended fraction of $12\%$ to $17\%$.

\subsubsection{Comparing with the Chandra plume}\label{subsubsection{chan_plume}}

For comparison with the Fe K map, we also show the soft X-ray Chandra contours that indicate the X-ray plume within the nuclear outflow \citep{2002MNRAS.335..241S}. 

The nuclear outflow resembles a hollow cone and was first identified via optical images and spectra \citep{1989PASJ...41.1107N, 1990ApJS...74..833H}.\ The structure is aligned with the galaxy minor axis \citep{1996A&A...308L...1M} with an opening angle of $\sim75^{\circ}$ \citep{1990ApJS...74..833H, 10.3389/fspas.2017.00046}.\ The NW lobe is tilted toward the observer and reaches a distance of $\sim1.8$ kpc from the nucleus \citep{10.3389/fspas.2017.00046}.\ The walls of the hollow cone of the ionized gas observed in the optical are filled with X-ray emitting gas as an edge-brightened plume that extends to at least 500 pc to the northwest mapped with Chandra and XMM \citep{2002MNRAS.335..241S}.\ The size of $\sim1-2$ kpc for the nuclear outflow is significantly smaller than the total Fe K nebulosity uncovered in the present work.

\subsection{Spectral analysis \label{sec:spectralanalysis}}

\begin{figure*}[ht!]
\includegraphics[width=0.4\linewidth]{ngc4945_diffuseFeK.png}
\includegraphics[width=0.5\linewidth]{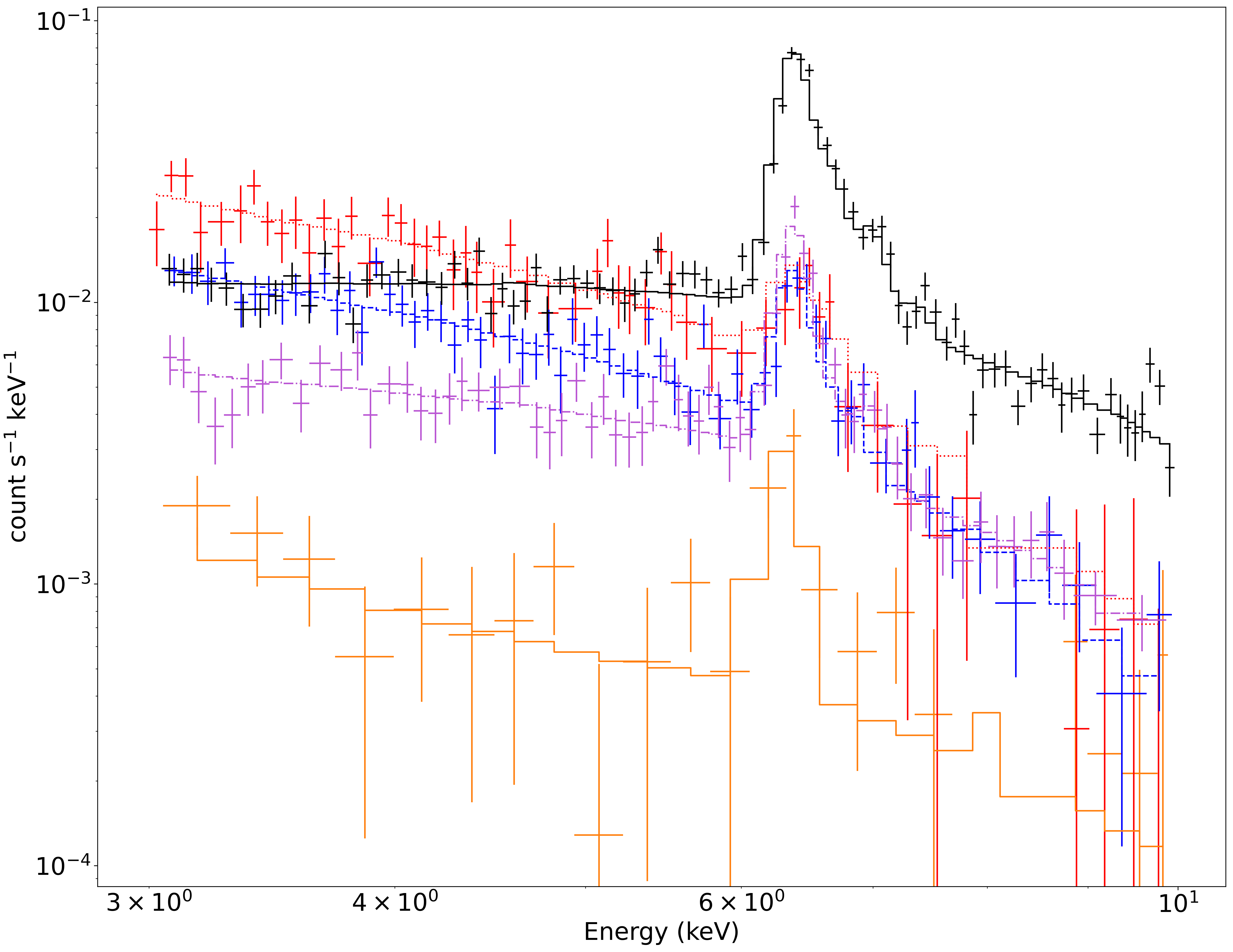}\\
\caption{\footnotesize{{\bf Left:} Fe K$\alpha$ map with contours overlaid (green). The lowest contour level is drawn at $2\sigma$ above the source-free background. The extraction regions used to derive the radial profile and for spectral extraction are the circles in black solid (radius = $16''$), mauve dot-dashed (radius = $32.5''$), blue dashed (radius = $65''$), and red dotted (radius = $130''$). The orange polygon is the ``hook'' extraction region.  {\bf Right:} The 3-10 keV spectra corresponding to the spatial extraction regions in Figure \ref{fig:fig5}. Spectral results are discussed in the text.}
\label{fig:fig7}}
\end{figure*}

The spectra provide complementary information to the imaging analysis. We first examined the Chandra spectra of the brightest point sources within the nebulosity region in the Fe K map (Figures  \ref{fig:fig2}, \ref{fig:fig4}, \ref{fig:fig5}).\ All are hard X-ray point sources but none possess significant Fe K$\alpha$ counts in their Chandra spectrum.\ We proceed under the assumption that the Chandra point sources within this region are not, individually, significant contributors to Fe K$\alpha$ emission; although they obviously contribute to the continuum levels.\ Additionally, the faintness of some of these sources above $\sim5$ keV in Chandra makes the hard X-ray fluxes difficult to constrain from spectral fitting.\ A direct measurement of the ensemble of individual Chandra sources via spectral fitting will have substantial error.\ Instead, we focus our spectroscopic analysis on the EPIC pn.

For the pn spectra, we use nuclear and annular regions with extraction radii of $16''$ (annulus 0 - A0), $16'' - 32.5''$ (annulus 1 - A1), $32.5'' - 65''$ (annulus 2 - A2), and $65'' - 130''$ (annulus 3 - A3), respectively (Figures \ref{fig:fig4}, \ref{fig:fig7}).\ We extracted spectra as described in Section 2, and only fit the spectra from 3 to 10 keV. We chose simple continuum models to apply to all spectra: model 1 (M1) is an absorbed power law adding Gaussians to model the emission lines, model 2 (M2) is an absorbed APEC model adding Gaussians to model the non-thermal emission lines.\ Model 3 (M3) is a combination of both M1 and M2. These models are not necessarily meant to be physical descriptions of the continuum, but rather provide a baseline against which to measure the hard X-ray line emission.\

All of the spectra and their best fitting models are shown together in Figure \ref{fig:fig7}.\ To generate this plot for illustration purposes, the spectra were all fitted together in {\tt xspec} with the line energies and line widths fixed to be identical.\ M3 was used to allow for different continuum shapes as per the below discussion of the individual fits.\ This figure is presented to illustrate a direct comparison of the Fe I K$\alpha$ line, which is detected with high significance to a radius of more than $200''$ away from the AGN. 

\subsubsection{pn spectra in the central $\le130''$ region\label{subsubsec:annular_spectra}}

\begin{figure*}[ht!]
\includegraphics[width=0.99\linewidth]{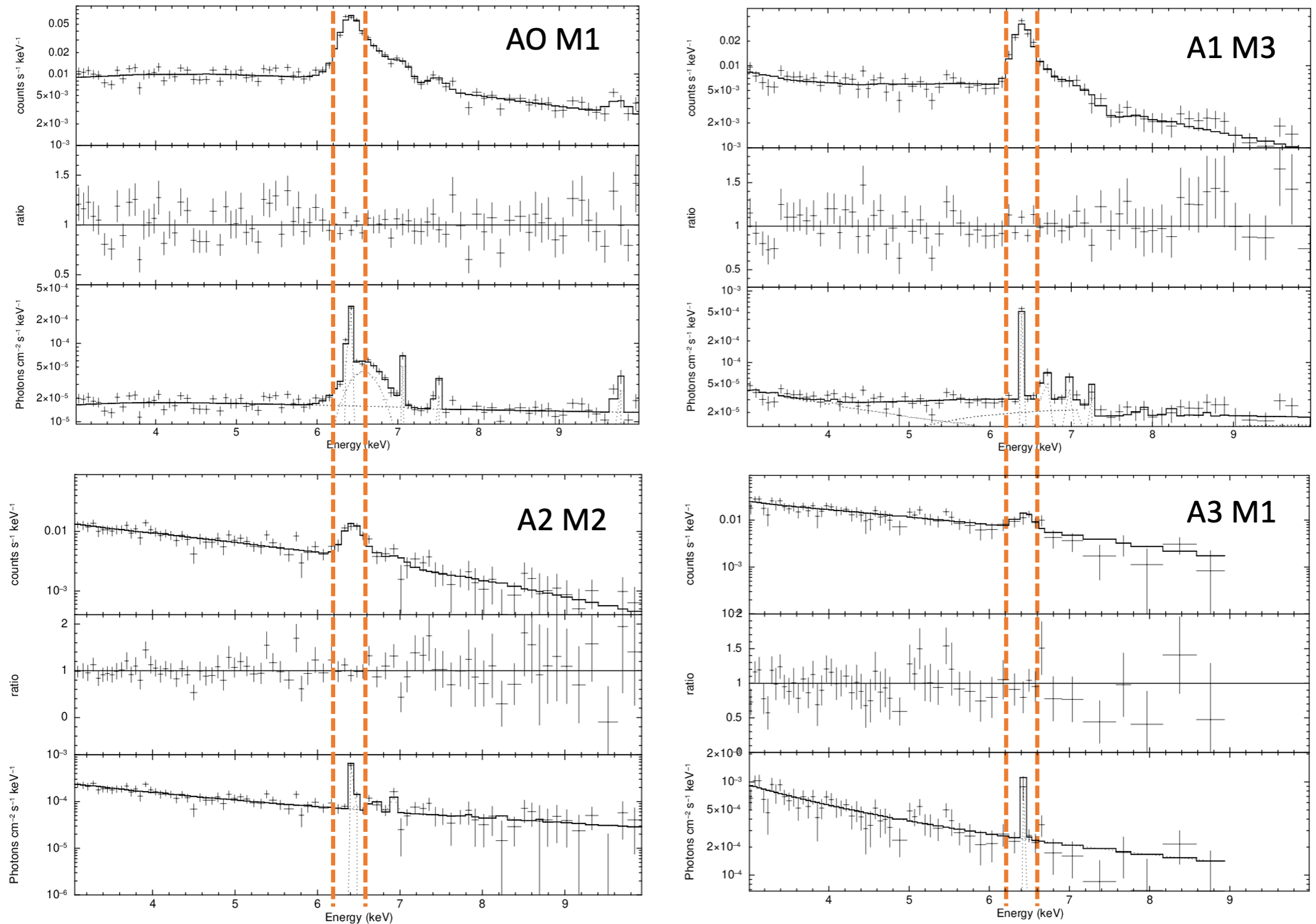}
\caption{\footnotesize{EPIC pn spectra and models listed in Table \ref{tab:spectral}.\ The top, middle and bottom panel within each quadrant is the data (plus model), the ratio of the data to the model and the unfolded spectrum.\ The dashed vertical lines represent the 6.2 to 6.58 keV bandpass used for the Fe K images.\ Top: A0 and A1 spectra fitted with models M1 and M3, respectively.\ For AO, we show the best-fitting power law adding five Gaussians.\ The line at $\sim9.7$ keV is not listed in the table but is likely a residual background feature.\ For A1, model M1 (power law with three Gaussians) does provide a better fit than M3, but we show M3 to illustrate the relative strength of the hot thermal component assuming solar abundances.\ Bottom: A2 and A3 spectra fitted with models M2 and M1, respectively.}
}
\label{fig:fig8}
\end{figure*}

We now discuss the individual spectral fits for A0 through A3.\
The spatially distinct pn spectra and select models are shown in Figure \ref{fig:fig8}.\ The best-fitting parameters from all model fits are listed in Table~\ref{tab:spectral}.

A0 contains the AGN core.\ Due to the strong non-thermal AGN component, M2 provides a poor fit.\ We therefore use M1 as the baseline model.\ The data require four Gaussians: the Fe K$\alpha$ line, Fe K$\beta$, Ni K$\alpha$, and a $\sim6.6$ keV Fe line from collisionally-ionized gas.\ The feature is broad, suggesting a complex of unresolved lines.\ There is also a line at 9.7 keV, which is likely a residual background feature.\ Our best fit is derived for M1 with the absorption and photon index allowed to vary (see Figure \ref{fig:fig8}).\ We also tried adding a partially-covered absorber and fixing the photon index at 1.7.\ This revealed a high column density of $5.2\times10^{23}$ cm$^{-2}$, but produced a slightly worse fit.\  We do note that this derived column density is identical to N(H2) = $\sim5\times10^{23}$ cm$^{-2}$ estimated via IR SED fitting \citep{2020yCat..36420166B}.

There is some ambiguity in the A0 spectrum regarding the Fe K lines.\ The best-fit $\sim6.6$ keV Gaussian line prefers to be broad.\ If we use M1 without extra absorption and also allow $\Gamma$ to vary while fixing this line energy at 6.7 keV and the width at 0.01 keV, the EW is $140\pm20$ eV.\ If we then allow the 6.4 keV line to be broadened, we find $\sigma=0.05\pm0.02$ and the 6.4 keV EW is $\sim1$ keV.\ Interestingly, allowing the $\sim6.7$ keV line to be the broad Fe K feature results in $\Delta\chi^2$ = 10 over a narrow line, so this is a significant improvement in the fit.\ It isn't possible however to constrain the energy of an additional narrow feature so the data do not require a third Fe K line.\ We conclude that there is obvious complexity in this region of the spectrum for A0.\ Without a more detailed analysis that is beyond the scope of this paper we cannot determine if this is due line emission, line broadening, or a complex continuum.

We next tried M3 and added a second absorbing medium with the assumption that the thermal component is located physically outside the clouds that obscure the heavily absorbed AGN continuum.\ We chose an APEC iron abundance of solar to be consistent with prior studies \citep{2002MNRAS.335..241S}.\ The photon index was fixed at $\Gamma = 1.7$ and we derive a large nuclear column density of $94\times10^{23}$ cm$^{-2}$.\ The second absorber has a column density of $7\times10^{22}$ cm$^{-2}$, which is consistent with the largest observed columns in the galaxy disk from H~I maps \citep{Ianjamasimanana2022}.\ This model also yields a larger normalization for the power law than M1 and the Fe K$\alpha$ line has an inferred line luminosity of $7.2 \times 10^{38}$ erg\,s$^{-1}$ and an EW of 800 eV.\ 

For A1, we obtain an excellent fit with M3, and the gas temperature is $8.2^{+3.7}_{-2.2}$ keV.\ The spectrum of A1 contains the wings of the PSF and so we expect about 40\% of the AGN counts to be scattered into this region.\ But the Fe K line EW increases, which likely indicates the relatively larger presence of the extended, circumnuclear Fe K in this region \citep{2012MNRAS.423L...6M} and Section \ref{subsubsec:radial}. 


\begin{splitdeluxetable*}{lccccccBccccccccBccccccc}
\tabletypesize{\scriptsize}
\tablewidth{0pt} 
\tablecaption{EPIC pn Spectral Fitting Results for 3 to 10 keV \label{tab:spectral}}
\tablehead{
\colhead{Spectrum|Model} & \colhead{N$_{H}$} & \colhead{kT} & \colhead{$\Gamma$} & \colhead{Norm$_{\Gamma}$} & \colhead{E$_{1}$(Fe I K$\alpha$)} & \colhead{$\sigma_{1}$} & \colhead{Norm$_{1}$} & \colhead{Line Flux$_{1}$} & \colhead{Lumin$_{1}$} & \colhead{EW$_{1}$} & \colhead{E$_{2}$(Fe XXV)} & \colhead{$\sigma_{2}$} & \colhead{Norm$_{2}$} & \colhead{EW$_{2}$} & \colhead{E$_{3}$(Fe K$\beta$)} & \colhead{Norm$_{3}$}  & \colhead{EW$_{3}$} & \colhead{E$_{4}$(Ni K$\alpha$)} & \colhead{Norm$_{4}$} & \colhead{EW$_{4}$(Ni K$\alpha$)} & \colhead{$\chi^{2}_{\nu}$} \\
\colhead{} & \colhead{$\times10^{22}$ cm$^{-2}$} & \colhead{(keV)} & \colhead{} & \colhead{x$10^{-4}$} & \colhead{(keV)} & \colhead{(keV)} & \colhead{x$10^{-5}$} & \colhead{x$10^{-13}$ erg\,cm$^{-2}$\,s$^{-1}$} & \colhead{x$10^{38}$ erg\,s$^{-1}$} & \colhead{(keV)} & \colhead{(keV)} & \colhead{(keV)} & \colhead{x$10^{-5}$} & \colhead{(keV)} & \colhead{(keV)} & \colhead{x$10^{-5}$}  & \colhead{(keV)} & \colhead{(keV)} & \colhead{x$10^{-5}$} & \colhead{(keV)} & \colhead{} \\
} 
\colnumbers
\startdata
 A0|M1  & $2.5^{+2.0}_{-1.9}$ & -- &  $0.5^{+0.2}_{-0.2}$ & $0.4^{+0.2}_{-0.1}$ & $6.41^{+0.02}_{-0.01}$  & $0.02^{+0.03}_{-0.02}$ & $2.03^{+0.53}_{-0.35}$ & 2.2 & 5.1 & $0.47^{+0.34}_{-0.08}$ & $6.59^{+0.15}_{-0.09}$  & $0.21^{+0.05}_{-0.10}$ & $2.28^{+0.33}_{-0.41}$ & $0.49^{+0.08}_{-0.14}$ & $7.06^{+0.04}_{-0.04}$ & $0.36^{+0.16}_{-0.10}$ & $0.15^{+0.07}_{-0.04}$ & $7.48^{+0.07}_{-0.05}$ & $0.20^{+0.09}_{-0.04}$ & $0.14^{+0.07}_{-0.06}$ &  1.29 \\
 A0|M1(pc)$^a$ & 0.14f / $52^{+9}_{-10}$ & --  &  1.7f & $10.2^{+0.3}_{-0.2}$ & $6.41^{+0.01}_{-0.02}$ & 0.01f & $2.45^{+0.41}_{-0.54}$ & 2.4 & 5.6  & $0.68^{+0.08}_{-0.10}$ &  $6.63^{+0.07}_{-0.06}$ & $0.13^{+0.06}_{-0.09}$ & $1.30^{+0.60}_{-0.48}$ & $0.26^{+0.10}_{-0.10}$ & $7.07^{+0.05}_{-0.04}$ & $0.39^{+0.12}_{-0.13}$ & $0.18^{+0.05}_{-0.06}$ & $7.48^{+0.05}_{-0.05}$ & $0.29^{+0.08}_{-0.08}$ & $0.22^{+0.07}_{-0.06}$ & 1.32  \\
 A0|M3(2N$_{\rm H}$)$^{b,c}$ & $7.4^{+1.5}_{-1.6}$ / $94^{+16}_{-24}$ & $6.9^{+1.0}_{-1.0}$ & 1.7f & $25.9^{+12.5}_{-7.2}$ & $6.42^{+0.01}_{-0.02}$ & 0.01f & $3.09^{+0.21}_{-0.20}$ & 3.1 & 7.2 & $0.80^{+0.06}_{-0.05}$ & -- & -- & -- & -- & $7.10^{+0.07}_{-0.06}$ & $0.24^{+0.16}_{-0.11}$ & $0.10^{+0.07}_{-0.04}$ &  $7.44^{+0.05}_{-0.05}$ & $0.42^{+0.10}_{-0.10}$ & $0.30^{+0.08}_{-0.08}$ &  1.30 \\ 
 A1|M1  & $0.5^{+2.3}_{-0.5}$ & -- & $0.62^{+0.15}_{-0.26}$ & $7.85^{+5.55}_{-2.00}$ & $6.40 ^{+0.02}_{-0.02}$ & 0.01f & $3.7^{+0.5}_{-0.4}$ & 3.7 & 8.6 & $1.00^{+0.17}_{-0.14}$ & $6.70^{+0.06}_{-0.11}$ & 0.1f & $1.15^{+0.40}_{-0.31}$ & $0.14^{+0.05}_{-0.04}$ & $7.06^{+0.07}_{-0.07}$ & $0.54^{+0.20}_{-0.19}$& $0.18^{+0.06}_{-0.07}$ & -- & -- & -- & 0.90 \\
 A1|M3$^b$  & 0.14f / $53^{+18}_{-15}$ & $8.2^{+3.7}_{-2.2}$  & 1.7f & $12.4^{+5.1}_{-3.2}$ & $6.40^{+0.02}_{-0.01}$ & 0.01f  & $3.7^{+0.04}_{-0.3}$ & 3.7 & 8.6 & $0.93^{+0.08}_{-0.09}$ & -- & -- & -- & -- & $7.2^{+0.2}_{-0.2}$ & $0.32^{+0.25}_{-0.19}$ & $0.12^{+0.06}_{-0.07}$ & -- & -- & -- & 1.00  \\
 A2|M1  & 0.14f &  --  & $1.65^{+0.18}_{-0.17}$  & $15.5^{+4.5}_{-3.5}$  & 6.4f & 0.01f & $3.60^{+1.07}_{-1.70}$ & 3.7  & 8.6 & $0.40^{+0.10}_{-0.20}$ & $6.57^{+0.08}_{-0.09}$ & 0.01f & $1.80^{+1.26}_{-0.81}$ & $0.15^{+0.1}_{-0.07}$ & -- & -- & -- & -- &-- &-- & 0.79 \\
 A2|M2$^d$  & 0.14f  &  $11.5^{+9.7}_{-3.0}$   &  --  & -- & $6.43^{+0.03}_{-0.03}$ & 0.01f & $4.7^{+0.8}_{-0.8}$ & 4.8 & 11.1 & $0.60^{+0.10}_{-0.11}$  & --  & -- & -- & -- & -- & -- & -- & -- & -- &-- & 0.78 \\
 A3|M1(nG)$^e$ & 0.14f  &  --  &  $1.60^{+0.20}_{-0.14}$  & $53.0^{+41.0}_{-75.0}$ & -- & -- & -- & --  &  --  & --  & -- & -- & -- & --  & -- & -- & -- & -- &-- &-- & 1.15  \\
 A3|M1  & 0.14f  &  --  &  $1.77^{+0.23}_{-0.23}$  & $69.2^{+24.0}_{-19.2}$ & $6.45^{+0.07}_{-0.07}$ & 0.01f & $7.20^{+2.80}_{-2.91}$ & 7.4 & 17.1 & $0.30^{+0.12}_{-0.12}$ &  6.7f  & 0.01f & $1.3^{+3.3}_{-1.3}$  & $0.04^{+0.10}_{-0.04}$   & -- & -- & -- & -- &-- &-- & 0.88  \\
 A3|M2$^f$  & 0.14f  & $9.4^{+8.6}_{-3.2}$  &  --  & -- & $6.42^{+0.07}_{-0.06}$  & 0.01f & $7.01^{+2.81}_{-2.93}$ & 7.2  & 16.7 & $0.24^{+0.11}_{-0.09}$ &  --  & --  & --  & --   & -- & -- & -- & -- &-- &-- & 0.95  \\
Hook|M1 & 0.14f  & --  & $1.72^{+1.54}_{-1.14}$ & $0.46^{+1.50}_{-0.38}$  & $6.35^{+0.05}_{-0.04}$ & 0.01f & $0.41^{+0.14}_{-0.17}$ & 0.42 & 0.97 & $2.1^{+0.7}_{-0.6}$ & --  & --   & --  & --  & --   & -- & -- &--  & -- &-- & 1.03 \\
\enddata
\tablecomments{Model 1 (M1) is an absorbed power law, model 2 (M2) is an absorbed APEC model, model 3 (M3) is a combination of M1 and M2.\ The power law normalization is in units of photons keV$^{-1}$ cm$^{-2}$ s$^{-1}$ at 1 keV.\ The Gaussian normalization is quoted as source frame photons cm$^{-2}$ s$^{-1}$.\ $^a$This model includes a partially covered power law continuum with the covering fraction = 0.87(0.84-0.89).\ $^b$A0M3 and A1M3 APEC abundance is fixed at solar.\ $^c$This model contains a second absorption component intrinsic to the AGN (i.e., the torus) to test the value that would be derived from the data if only the AGN is assumed to be heavily absorbed. The derived value is N$_{\rm H} = 94\pm{20} \times10^{22}$ cm$^{-2}$.\ $^d$Abundances are derived as 0.34 (0.11-0.60) solar.\ $^e$nG=No Gaussians included.\ $^f$Abundance is fixed at 0.3 solar. All quoted errors are at the $90\%$ confidence level for one free parameter.}
\end{splitdeluxetable*}  

For A2, M1 and M2 yield equally good fits.\ The Fe I K$\alpha$ line is clearly detected and the $\sim6.6$ keV line is detected with moderate significance.\ For M1, the data allow absorption up to $4\times10^{22}$ cm$^{-2}$.\ If we use a single Gaussian to model the 6.4 keV line and allow it to be broad with no other Gaussian, we have $\sigma = 0.08^{+0.04}_{-0.05}$ keV and EW $\sim 750$ eV.\ Adding a second line is marginally significant but the feature also tends to be broad, implying line blending.\ If we instead make the 6.4 keV line narrow, we obtain EW(Fe K$\alpha$) $\sim640$ eV.\ Now adding the second Gaussian gives $\Delta\chi^2$ = 5 for the addition of two free parameters. If we fix the energy of the first line at 6.4 keV, we derive a line energy of $6.57^{+0.08}_{-0.09}$ keV (Table \ref{tab:spectral}).\ The errors in line energy for the second feature are larger when we allow the 6.4 keV line energy to vary. 

Examining the line blending we tried a narrow and broad 6.4 keV line in deriving a self-consistent abundances.\
For A2 M2, the best-fitting thermal temperature is $11.5^{+9.7}_{-3.0}$ keV and if we allow the Fe K 6.4 keV line to be broad ($\sigma\sim0.06$ keV) the APEC model Fe abundance is 0.25$\pm0.25$.\ If we fix the Fe abundance at the solar value, the reduced $\chi^2$ is acceptable at 0.88 but kT increases to $24^{+14}_{-8}$ keV.\ Allowing the abundance to be free and with a narrow Gaussian fixed at 6.4 keV, yields a reduced $\chi^2$ of 0.78 and Z=$0.34^{+0.26}_{-0.23}$ solar.\ The significance for solar vs. non-solar abundances for the APEC model is $\Delta\chi^2$ = 7.\ Allowing the features to be broad significantly affects the fit by driving the temperature to a larger value that is inconsistent with the A0 and A1 measurements.\ We therefore prefer using narrow lines, but in either case, the abundances are less than solar.\ A reduced abundance in the outer regions is consistent with abundance measurements in other starburst galaxies that are known to change with distance from the galaxy center \citep[NGC 253,][]{2011ApJ...742L..31M}.\ Regardless of the model, the EW of the Fe I K$\alpha$ line is still quite high in A2 - $\sim400 - 600$ eV. 

For A3, we initially tried M1 without a Gaussian (nG in Table \ref{tab:spectral}) to test the significance of any line emission.\ The fit was marginally acceptable but adding a Gaussian at 6.4 keV dramatically dropped $\chi^2$ by 19 and in this case the EW is $300\pm120$ eV.\ An APEC model with abundances fixed at 0.3 solar (as per A2 M2 above) provides a slightly worse fit, due to the Fe feature in the model being predicted as too strong.\ The gas temperature is constrained at $9.4^{+8.6}_{-3.2}$ keV.\ For this model the 6.4 keV line EW is 240 (+100, -90) eV and the upper limit on the EW of a 6.7 keV line is 140 eV.

We conclude from the spatially resolved spectra within $130''$ of the nucleus that the neutral Fe K$\alpha$ line remains strong out to $130''$, and the line $L_{\rm{X}}$ ranges from $\sim5.1\times10^{38}$ to $\sim1.7\times10^{39}$ ergs sec$^{-1}$.\ There is line blending in the region of the iron emission lines that provides uncertainty in the exact line strengths and abundances derived from the $\sim6.6-6.7$ keV feature, but the data prefer a scenario that is consistent with solar abundances in the galaxy center and dropping below solar further out.\ The derived gas temperatures within A1, A2 and A3 are consistent with each other as well as the 2001 XMM observation of NGC 4945 \citep{2002MNRAS.335..241S} and BeppoSAX \citep{2000A&A...356..463G}. The Fe I K$\alpha$ line emission is measured with normalizations of $2 - 3 \times10^{-5}$ photons cm$^{-2}$ s$^{-1}$ (AO), $3.7\times10^{-5}$ photons cm$^{-2}$ s$^{-1}$ (A1), $3.6 - 4.7 \times10^{-5}$ photons cm$^{-2}$ s$^{-1}$ (A2), and $7.2\times10^{-5}$ photons cm$^{-2}$ s$^{-1}$ (A3).\ We also note that BeppoSAX finds a line normalization of $\sim4\times10^{-5}$ photons cm$^{-2}$ s$^{-1}$, which is consistent with our nuclear measurement at $<32.5''$ (A0 + A1) \citep{2000A&A...356..463G}.

The Fe K line luminosity in A3 is particularly large: $\sim1.7\times10^{39}$ ergs sec$^{-1}$.\ We point out again that all point sources are included in this spectrum and going back to our examination of the Chandra point sources (Table \ref{tab:pointsources}), 6 out of 10 of the brightest point sources are contributing to the A3 region.\ These contribute $\sim10\%$ to the total flux compared with the AGN.\ It is possible that a similar fraction is contributed to the Fe K emission from individual sources that are X-ray binaries or ULX-like in nature.\ For example, source \#7 in A3 is a variable ULX \citep{1996MNRAS.281L..41B}.\ We discuss further the possible level of contribution from stellar sources in Section \ref{subsubsection:stellar}.

\subsubsection{Scales beyond 130 arcseconds - the hook region\label{subsubsec:hook}}

For the ``hook'', we used a polygon region to trace the visible extended emission (Figure \ref{fig:fig7}).\ In this faint region, we carefully examined the effects of the pn detector background, which includes spectral features of fluorescence from the detectors and surrounding structure \citep{2002A&A...389...93L}.\ The most prominent in our case is the complex around 8 keV, due to Cu-K$\alpha$, Ni-K$\alpha$ and Zn-K$\alpha$.\ We restricted the data and background region extraction to the Cu hole where the complex is virtually absent (see Figure \ref{fig:fig1} - hard X-ray map), but even in this case, weak features are present in the background spectrum and source spectrum, at slightly different intensities due to the different spectral extraction locations.\ We verified by examining multiple background regions that the background subtraction is adequate to account for these features.

\begin{figure}[h]
\plotone{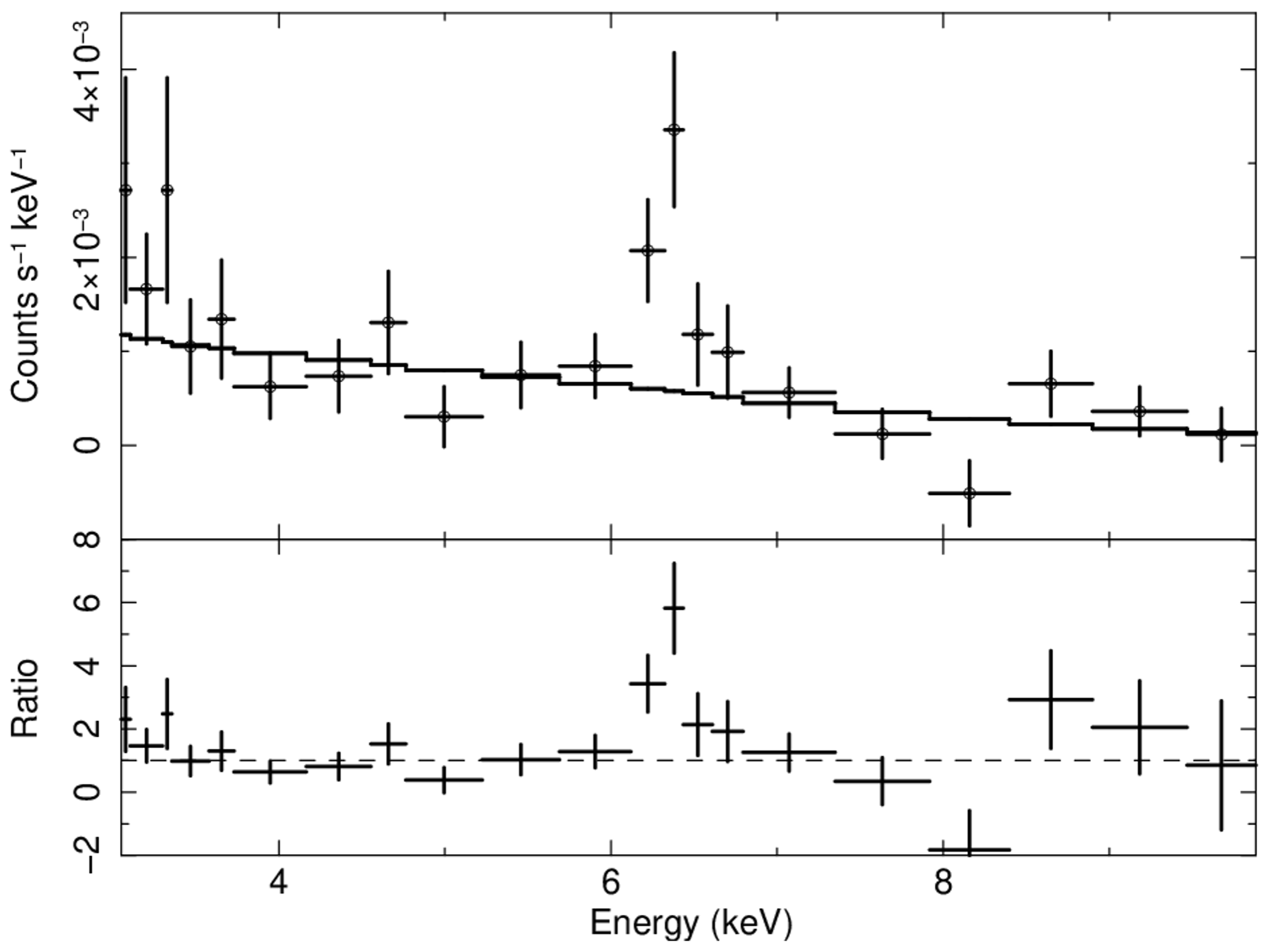}\\
\plotone{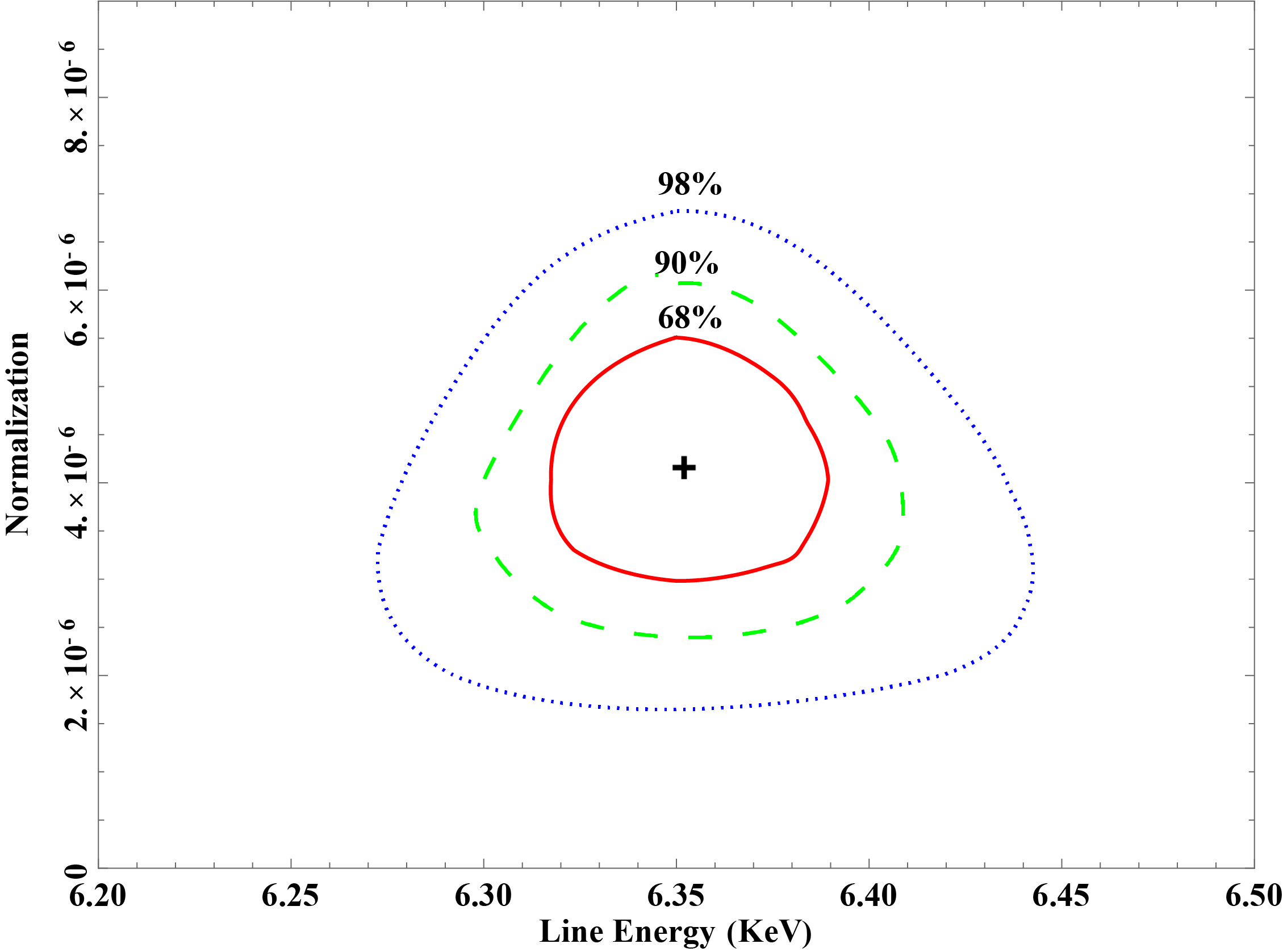}
\caption{\footnotesize{{\bf Top:} The 3 to 10 keV pn spectrum for the ``hook'' region fitted to an absorbed power law continuum model ($\Gamma\sim0.9$).\ {\bf Bottom:} Contours for line energy (keV) vs.\ line normalization (photons cm$^{-2}$ s$^{-1}$) at 68\%, 90\% and 98\% confidence for the fit when adding the Gaussian - see Table \ref{tab:spectral}. }
\label{fig:fig9}}
\end{figure}

We first used M1 (power law) with Galactic $N_{\rm{H}}$ and modeling no line emission.\ The fit is poor and yields $\Gamma=0.89 ^{+0.03}_{-1.88}$ with $\chi^2$/$\nu=1.3$.\ The residuals at $\sim6.4$ keV are striking (Figure \ref{fig:fig9}).\ Adding a narrow Gaussian yields $\Delta\chi^2$ = 24 for two free parameters (improving the fit at $>99$\% confidence) and $\chi^2$/$\nu$ = 1.03 (Table \ref{tab:spectral}).
 
In this case, the Fe K line EW is measured at $\sim2$ keV.\ The 2-10 keV flux in the hook is $2.3\times10^{-13}$\,erg\,cm$^{-2}$\,s$^{-1}$, which is about 12\% of the 2-10 keV AGN (observed) flux of $\sim1.9\times10^{-12}$ ergs cm$^{-2}$ s$^{-1}$.\ The Fe K line normalization is $4\times10^{-6}$ photons cm$^{-2}$ s$^{-1}$ with a flux of $\sim5\times10^{-14}$\,erg\,cm$^{-2}$\,s$^{-1}$, or roughly 2-3\% of this AGN continuum level.\ Adding absorption in addition to the Galactic value allows for a column density as high as $9\times10^{22}$ cm$^{-2}$ along the line of sight, consistent with the highest columns measured in the disk via other means.\ 

\begin{deluxetable*}{lcccccccc}
\tabletypesize{\scriptsize}
\tablewidth{0pt} 
\tablecaption{Extended fraction for Fe I K$\alpha$ for different techniques\label{tab:fraction}}
\tablehead{
\colhead{Data} & \colhead{Technique}& \colhead{Region} & \colhead{Point sources} & \colhead{Continuum} & \colhead{Background} & \colhead{Total} & \colhead{Comparison} & \colhead{Extended} \\
\colhead{} & \colhead{} & \colhead{}  & \colhead{Included?} & \colhead{Included?} & \colhead{Included?} & \colhead{} & \colhead{} & \colhead{Fraction} \\
} 
\colnumbers
\startdata
MOS1 & Radial profile & $16 - 130''$ & Y & Y & N & $0.84\pm0.25^a$ & 0.51 (PSF) & $39\pm12\%$ \\
pn & Radial profile & $16 - 130''$ & Y & Y & N &$1.36\pm0.16^a$ & 0.94 (PSF)  &  $31\pm3\%$ \\
Chandra & Encircled counts & $16 - 130''$ & N & Y & N & --  & -- & $\sim40\%$ \\
MOS1 & Radial profile & $32.5 - 130''$ & Y & Y & N  & $0.4\pm0.14^a$ & 0.22 (PSF) & $45\pm15\%$ \\
pn & Radial profile & $32.5 - 130''$ & Y & Y & N & $0.57\pm0.09^a$ & 0.35 (PSF) & $39\pm6\%$\\
pn spectra & modeling & $65 - 130''$ & Y & N & N & 18.2$^b$ & 7.7$^b$  & $\sim42\pm15\%$ \\
pn spectra & modeling & hook & N & N & N & 18.2$^b$ & 0.4$^b$ & $2.3\pm0.9\%$ \\
pn image & PSF-subtracted & $>32.5''$ & N & partial & Y & 4295$^c$ & 1560$^c$  & $36\pm4\%$ \\
pn image & PSF-subtracted & $>65''$ & N & partial & Y & 4295$^c$ & 1375$^c$ & $30\pm5\%$ \\
pn image & PSF-subtracted & $>32.5''$ & N & partial & N & 3271$^c$ & 564$^c$ & $17\pm5\%$ \\
pn image & PSF-subtracted & $>65''$ & N & partial & N & 3271$^c$ & 379$^c$ &  $12\pm5\%$ \\
\enddata
\tablecomments{$^a$Units are counts arcsec$^{-2}$.\ $^b$Values are line flux in units of $10^{-13}$ ergs cm$^{-2}$ s$^{-1}$.\ $^c$ Units are exposure-corrected counts.\ Quoted errors are statistical errors only.\ The most reliable measurements isolating the line emision are in bold.}
\end{deluxetable*}

\subsection{Analysis summary}
The results of our analysis indicate significant neutral Fe K$\alpha$ emission on scales as large as $\sim250''$ south of the nucleus.\ We summarize our results in Table \ref{tab:fraction}; the table compares the EPIC radial profiles, Chandra counts, the PSF-subtracted Fe K map, and line fluxes from spatially-resolved spectra.\ In each case we derive an extended fraction over comparable spatial scales and list whether point sources, continuum emission, and/or background are included in the estimate.\ 

From the spectral analysis, the fraction of Fe I K$\alpha$ is large comparing the annular regions to the nucleus.\ Averaging the line normalizations from the fits in \ref{tab:spectral}, we have a total line flux of $\sim18.2\times10^{-13}$ ergs cm$^{-2}$ s$^{-1}$.\ Adding A2 and A3 yields $\sim11\times10^{-13}$ ergs cm$^{-2}$ s$^{-1}$, or over 50\% of this total.\ But all pn point sources are included in the spectra and so this implies some point-source contribution for A2 and A3 to the Fe K line (Section \ref{subsubsec:annular_spectra}, Table \ref{tab:pointsources}).\ If we examine only the largest regions $>65''$ ($>1.3$ kpc), then we are comparing $\sim7.7\times10^{-13}$ ergs cm$^{-2}$ s$^{-1}$ (A3 plus hook) Fe K plus continuum with the total, which equates to the extended emission at $\sim$kpc scales making up $\sim42\%$ of the Fe I K$\alpha$ plus continuum emission in the galaxy.\ If we only consider the hook region, the measured line flux equates to $\sim2.3\%$ of the total background-subtracted Fe K emission.

The images are more straightforward to interpret than the spectra.\ There is no ideal case where we are able to completely isolate the Fe K line emission on its own, but the exposure-corrected PSF-subtracted map (and background subtracted) provides the most reliable estimate of the Fe K$\alpha$ extended fraction at large scales being from $12-17\%$.\ We have highlighted in bold in Table \ref{tab:fraction} the three most reliable estimates and the regions these cover.\ We conclude that for emission on scales lager than $32.5''$ ($>680$ pc) we estimate the approximate extended fraction for Fe I K$\alpha$ to be $\sim15\%$ in NGC 4945.

\section{Discussion\label{sec:discussion}}

We have found extended Fe I K$\alpha$ emission on scales as large as 10 kpc in NGC 4945.\ This is above and beyond the circumnuclear Fe K emission extended on a $\sim200$ pc scale seen with Chandra, which we confirm.\ 
From radial profiles, the PSF-subtracted and background-subtracted pn image, and spatially resolved spectra, we estimate that $\sim15\%$ of the neutral Fe K$\alpha$ emission is extended beyond scales of 680 pc out to $\sim10$ kpc.\ The existence of extended hard X-ray emission at these scales is not surprising.\ The Medium Energy
Concentrator Spectrometer (MECS) instrument on BeppoSAX detected extended emission from 2-6 keV at least out to a $\sim4'$ radius \citep[the innermost 5 kpc;][]{2000A&A...356..463G} - including what seems to be a marginal detection between 6 and 7 keV (their Figure 5).\ The emission region also showed a slight elongation along the galaxy plane, which these authors suggested could signal a truly diffuse ISM.

The Fe I K$\alpha$ line traces regions of cool/cold dense gas and extended regions of this diagnostic feature are becoming a more commonly seen in nearby AGN.\ Studies with Chandra have now detected hard X-rays, including Fe K$\alpha$, from 100s of pc to kpc scales from the nucleus in Compton-thick (CT) AGN in galaxies such as Circinus \citep{kawamuro2019}, NGC 6240 \citep{2014ApJ...781...55W} and ESO 428-G014 \citep[e.g.][]{2017ApJ...842L...4F}. 
This appears directly related to the AGN activity in that the emission follows the ionization cone and cross-cone, and sometimes the jet direction (e.g., ESO 428–G014, \citealp{2018ApJ...855..131F, 2018ApJ...865...83F, 2019ApJ...870...69F}; NGC 7212, \citealp{2020ApJ...891..133J}). 
Hard X-ray emission on $\sim$kpc scales in the direction of the ionization cone is also seen in Circinus \citep{2014ApJ...791...81A} and NGC 1068 \citep{2015ApJ...812..116B}.\ We note that currently, extended Fe I K$\alpha$ has only been seen in Compton-thick (CT) AGN \citep{2020ApJ...900..164M}.\ The fact that we can pick up the faint extended hard X-ray emission in CT AGN may be a selection effect due to the difficulty of measuring such emission in the proximity of less obscured AGN.\ 

NGC 4945 differs from these other AGN in that it lacks a traditional narrow line region and [\ion{O}{3}] is not detected \citep{1986Ap&SS.121..403A}.\ On the other hand, it is a well known starburst galaxy \citep{1990ApJS...74..833H,2020ApJ...903...50E}.\ 
Evidence for Fe K$\alpha$ in extended regions is also found in traditional starbursts like M82 \citep{liu2014} and NGC 253 \citep{2011ApJ...742L..31M}, which have many similarities with NGC 4945 \citep{1990ApJS...74..833H}.\
The origins of the Fe I K$\alpha$ emission in M82 and NGC 253 appears to be consistent with their X-ray binary populations and fluorescence from molecular clouds in the starbursting regions.\ But given their similar star-forming rates and superwind (including soft X-ray) characteristics \citep{1990ApJS...74..833H}, it is important to examine NGC 4945 in the context of other starburst galaxies.

\subsection{What the images tell us}\label{subsec:image_disc}

The Fe I K$\alpha$ emission region extends $\sim480''$
along the plane of the galaxy, and to $\sim250''$ above the plane, while the overall extension to the NW and SE begins to exit the galaxy plane in projection and in a general direction angled at $\sim60^\circ$ with respect to the galaxy plane (Figure \ref{fig:fig6}).\ 
To the SE, this aligns with the direction where we see the largest soft X-ray emission extension with respect to the galaxy (Figure \ref{fig:fig1}).\ 
Although we do not discuss the soft X-ray emission directly in this work, it is likely due to the starburst activity \citep{2002MNRAS.335..241S}.\ The more compact soft X-ray nuclear outflow/plume perpendicular to the galaxy plane imaged with Chandra and XMM \citep{2002MNRAS.335..241S} is overlaid in Figure \ref{fig:fig6}.\ The nuclear outflow and the Fe K$\alpha$ nebulosity may or may not be related.\ We discuss possibilities in \ref{subsubsec:relic}.

\begin{figure}[h]
\plotone{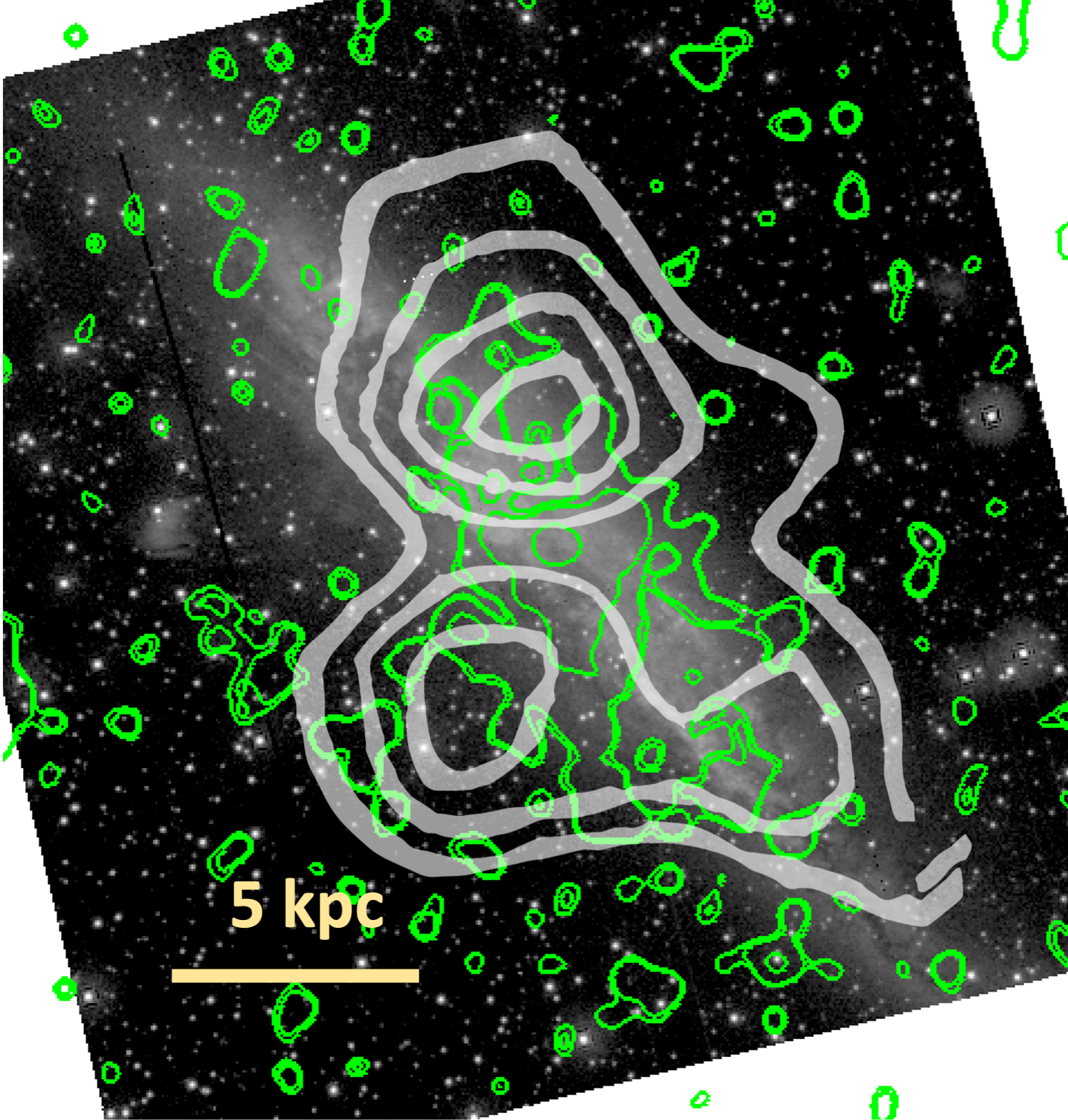}
\caption{\footnotesize{Contours of linearly polarized radio emission at 4.75 GHz \citep{1989A&A...216...39H} (white) and contours of the PSF-subtracted Fe K map (green, Figure \ref{fig:fig6}, right panel) overlaid on the OM B-band image.}
\label{fig:fig10}}
\end{figure}

An obvious source of Fe K$\alpha$ would be X-ray irradiation of molecular clouds and so we searched the literature for evidence of structures in the disk of the galaxy that might indicate the presence of these clouds.\ In the optical, the galaxy rotation curve is
linear, with features consistent with properties of other late-type Sd galaxies \citep{peterson1980}.\ There is previously undetected faint H~I emission extending far beyond the galaxy in a region of size 32 arcmin by 18 arcmin \citep[62 × 35 kpc;][]{Ianjamasimanana2022}.\ H~I measurements indicate a galaxy mass of $\sim1.4\times10^{11}$ solar masses within a radius of $380''$, along with evidence for a bar, which may be fueling the central starburst \citep{2001A&A...372..463O}.\ H~I emission from the inner regions at velocities of $>200$ km s$^{-1}$ compared to the systemic velocity are related to an outflow that is directed towards the halo \citep{2001A&A...372..463O}, and there are extensions on both sides of the galaxy’s major axis indicating that the kinematics in the halo do not follow the kinematics of the disk.\ The irregularity in the velocity/structure patterns of H~I are similar to that seen for superwinds in other starbursts \citep[for example,][]{2015MNRAS.450.3935L}, so the images are consistent with outflows that could be starburst-driven in nature.

Fe I K$\alpha$ emission with such large EWs as 2 keV (Table \ref{tab:spectral}) is not expected for a starburst (Section \ref{subsec:starburst}).\ Looking at specific deviations from these general kinematic trends in the atomic and molecular gas, there are irregularities that align with the orientation of the Fe K$\alpha$ extent.\ H~I maps and CO maps show unusual velocities in some areas of the galaxy disk and the halo, particularly toward the south-east \citep{dahlem1993, Ianjamasimanana2022} - which is in the direction of orientation of the Fe K nebulosity (blue arrow in Figure \ref{fig:fig6}).\ The question is whether AGN activity / AGN-driven outflows can produce the structures and patterns in the atomic and molecular gas maps.\ In CO and HCO+ channel maps, there are plumes whose morphology invokes molecular gas that has been ejected from the central regions \citep{2021ApJ...923...83B}. 

Taking this last point, NGC 4945 does show an unusual pattern of radio emission.\ Two maxima of linearly polarized radio emission are located at approximately 2.4 kpc out of the plane of the galaxy with the line joining them passing through the galaxy core at an angle of $40^\circ$ with respect to the z-axis \citep{1989A&A...216...39H}.\ These authors claim that the two distinct polarized sources suggest the presence of a strong ordered magnetic field at these locations, not just a general galaxy halo magnetic field.\ This high magnetic field alignment is likely to arise from a collimated outflow or infall of relativistic particles.\ Figure \ref{fig:fig10} shows a comparison of the radio polarization map with our Fe K map.\ The regions seem to align well overall.

If we are seeing evidence of past activity from the radio lobes, the central AGN activity would not need to be ongoing.\ We discuss in Section \ref{subsubsec:molecular} a way that may have allowed the AGN to have created large-scale filaments of gas clouds with enough column density and in the optimal orientation with respect to the X-ray continuum source to produce significant Fe K fluorescence.\ The alignment with the Fe K map could imply fluorescence almost tracing areas around the polarization maxima.\ We discuss the possibility for relic activity further in Section \ref{subsubsec:relic}.

\subsection{Starburst activity\label{subsec:starburst}}

NGC 4945 is a superwind galaxy \citep{1990ApJS...74..833H}, with clear evidence for an outflow that resembles a bubble-like structure seen in H$_2$ \citep{2000A&A...357...24M} and in the form of split optical emission lines with velocity separations ranging from $\sim300$ km s$^{-1}$ in the outer regions of the bubble to $\sim600$ km s$^{-1}$ near the center \citep{1990ApJS...74..833H}.\ The region of split emission lines extends to a distance of about 1 kpc along the galaxy minor axis and this region is also observed in soft X-rays with Chandra \citep{2002MNRAS.335..241S}.\ The starburst itself is concentrated within the central few hundred pc of the galaxy \citep{1990ApJS...74..833H, 2000A&A...357..898S} and the central AGN appears to be fully obscured in all directions \citep{2000A&A...357...24M}.\ \citet{2020ApJ...903...50E} estimate that less than 10\% of the ionizing photon luminosity of the AGN is escaping into the surrounding starburst region and thus the AGN would not, at present, be creating much (if any) ionized gas in the circumnuclear ISM. 

We measure a spatially-resolved hard X-ray component with a thermal temperature of kT $\sim10$ keV ($\sim1\times10^8$ K) out to possibly as far as $130''$ from the nucleus (Table \ref{tab:spectral}).\ The temperature of the gas is consistent with superwinds \citep{1985Natur.317...44C, 2005ARA&A..43..769V, 1990ApJS...74..833H} and our error bars on the gas temperature encompass either ionization from internal wind shocks or shocked ambient gas \citep{1990ApJS...74..833H}.\ For region A3, the 2 to 10 keV flux assuming a thermal continuum (M2) is $\sim2\times10^{-11}$ ergs cm$^{-2}$ s$^{-1}$ equating to a luminosity of $4.6\times10^{40}$ ergs s$^{-1}$.\ This consistent with X-ray luminosities for extended regions in other superwind galaxies  \citep[log L$_{\rm X}$ = $40-41.4$,][]{1990ApJS...74..833H}.

NGC 4945 has a current star formation rate of 11 M$_\odot$ yr$^{-1}$ \citep{1990ApJS...74..833H} to $\sim18$ $M_\odot$/yr \citep{Ianjamasimanana2022} - also similar to M82 and NGC 253.\ The current SFR is probably the most reliable way to estimate impact of the hot component.\ Helium-like iron emission is present at $\sim6.6 - 6.7$ keV with a normalization of $0.7\--2.3\times10^{-5}$ photons cm$^{-2}$ s$^{-1}$.\ The line EW appears to be strong in the galaxy center but drops toward the outer regions from $\sim500$ eV (A0) to $<140$ eV (A3).\ This is consistent with the most significant star formation occurring within the central $32.5''$ region of the galaxy.\ The hot gas that is present would be from the current/most recent generation of stars.\

The age of the actual starburst in NGC 4945 has been calculated at $>5$ Myr \citep{2000A&A...357..898S, 2000A&A...357...24M} and the older population of star clusters would have formed in an instantaneous burst of star formation roughly 5 Myr ago \citep{2020ApJ...903...50E}.\ This is consistent with the lack of strong He-like Fe K with distance from  the nucleus as this line is generated from young SNRs and XRBs.\ We take the advanced age of the starburst as a primary key to our understanding the large scale Fe K$\alpha$ emission in NGC 4945 as it suggests the age of a triggering event.


\subsection{Origins of the neutral Fe K$\alpha$ line}\label{subsec:fek_disc}

Fe I K$\alpha$ line emission is significantly detected in all of the pn spectra examined here that cover the extended Fe K nebulosity in NGC 4945.\ While the line is most significant at the location of the AGN, as expected, the line remains strong radially out to $\sim130''$ from the galaxy center.\ The EW ranges from $\sim500$ eV ($\le16''$) to $\sim1000$ eV ($16-32.5''$) to $\sim500$ eV ($32.5-65''$) to $\sim240$ eV ($65-130''$) to $\sim2000$ eV ($\sim130-220''$).\  We discuss here plausible scenarios for producing such significant Fe I K$\alpha$ in the galaxy disk. 

\subsubsection{{Ionization from stellar sources}\label{subsubsection:stellar}}

Along with obvious Fe lines of He-like emission (Fe XXV at 6.68 keV, Fe XXVI at 6.93 keV), Fe I K$\alpha$ has been detected in two other well known starburst galaxies - M82 \citep{liu2014, 2007ApJ...658..258S} and NGC 253 \citep{2011ApJ...742L..31M}.\ In M82, Fe I K$\alpha$ emission is seen on a scale of $\sim30$ arcsec \citep[0.6 kpc; ][]{liu2014} and the total luminosity of the line is $\sim4\times10^{37}$ ergs s$^{-1}$.\ 
These authors argue that the line luminosity is consistent with irradiation of dense molecular clouds by hard X-ray photons from point sources in the nuclear region of M82.\ In NGC 253, Fe K$\alpha$ emission is extended within the inner 60 arcsec$^{-2}$ region \citep{2011ApJ...742L..31M}.\ 
In the area with the strongest line fluxes in NGC 253, the luminosities of the Fe K$\alpha$ and Fe XXV lines are 1.2 and $2.7\times10^{37}$ ergs s$^{-1}$, respectively. The EWs are $\sim170$ eV and $\sim660$ eV, respectively.\ 
The best explanation for the line complex is a combination of tens to thousands of SNRs plus X-ray fluorescence from molecular clouds with a column density of $\sim10^{24}$ cm$^{-2}$, again irradiated by X-ray sources in the nuclear region of the galaxy.\ 
For M 82 and NGC 253, AGN luminosities are not required to produce the extended Fe K$\alpha$ emission. 

With NGC 4945 being a composite AGN and starburst galaxy, the first question is whether the Fe I K$\alpha$ features within the extended emission locations (the A1, A2, A3, hook regions) are due to stellar sources.\ As we showed in Section \ref{sec:imaging}, there are a few dozens of hard X-ray point sources embedded within the nebulosity.\ But even restricting ourselves to regions far away from the AGN - region A2 (L(line) $\sim10^{39}$ ergs s$^{-1}$, EW $\sim500$ eV), region A3 (L(line) $\sim2\times10^{39}$ ergs s$^{-1}$; EW $\sim240$ eV) and the hook (L(line) $\sim10^{38}$ ergs s$^{-1}$; EW $\sim2000$ eV) - the Fe I K$\alpha$ EWs and line luminosities are high.\ 

We have noted that with the line luminosities being so high in A2 and A3, and the line EW being the smallest in A3, individual bright point sources could contribute to the Fe K emission in these regions.\ To produce strong emission lines, other than individual ultraluminous X-ray sources (ULXs), point sources with the largest line luminosities are likely high-mass X-ray binaries (HMXBs) or black-hole X-ray binaries (BH XRBs).\ However, compared to other starbursts starbursts with similar star formation rates, NGC 4945 is NS-dominated with a steep X-ray luminosity function and a small population of $L_X > 10^{38}$ ergs s$^{-1}$ sources \citep{2018ApJ...864..150V}; there are relatively few ULXs and BH XRBs in NGC 4945.

Another clue that Fe I K$\alpha$ is not primarily from stellar sources is that while we may be seeing strong He-like iron emission in the central regions (see \ref{sec:spectralanalysis}), the line is weak or not significantly detected at larger radii.\ The lack of a 6.7 keV line implies a lack of star formation (i.e., SNRs, X-ray binaries).\ For comparison, the strong extended starburst in NGC 253 produces Fe XXV EWs up to $\sim660$ eV \citep{2011ApJ...742L..31M}.\ Non-eclipsing HMXBs have Fe K$\alpha$ line luminosities on the order of $\sim10^{34}$ \citep{2010ApJ...715..947T} and so $\sim10,000$ such sources would be required in NGC 4945 on the scale of a few kpc to produce even the emission in the hook (L(Fe K) $\sim10^{38}$ ergs s$^{-1}$).

We conclude that the extended Fe I K$\alpha$ cannot be due to an ensemble of X-ray point sources in these regions unless there are unseen populations of heavily obscured or eclipsing XRBs arranged within kpc-sized areas, which seems implausible.\ An alternative interpretation would involve ULXs or bright XRBs that have turned off.\ Three bright X-ray point sources observed with Chandra in 2014 are not detected in the pn image (Figures \ref{fig:fig4}, \ref{fig:fig5}, and Table \ref{tab:pointsources}) and perhaps these could have left behind residual scattering from surrounding molecular clouds.

\subsubsection{Reflection from molecular clouds caused by current / recent AGN activity}\label{subsubsec:molecular}

As we have stated the extended Fe I K$\alpha$ is likely from fluorescence from cold, dense molecular clouds in the ISM.\ For reflection from neutral material with significant optical depth we expect a low ionization Fe K line with equivalent width of about 1 keV, an Fe K edge feature, a Compton hump and a flat 2–10 keV continuum due to internal photo-absorption \citep{2010MNRAS.401..411Y,2009MNRAS.397.1549M}.\ The large EW of Fe K$\alpha$ in the ``hook'' of $2.1^{+0.7}_{-0.6}$ is particularly consistent with scattering interpretations of Fe K$\alpha$ lines \citep{2002ApJ...573L..81L}.\ Such large EWs occur from a photoionizing X-ray source that is not seen in the line of sight \citep[e.g.,][]{ 2018ApJ...853..146T}.\ But in these cases for AGNs, the lines were assumed to originate in the galaxy core and the fluorescence would occur within the molecular torus region with optically thick column densities of N$_{H}\ge10^{24}$ cm$^{-2}$.\ 


Overall, the reflected spectrum will depend upon the cloud density, ionization parameter, the ionizing spectrum, and any self absorption from intervening material \citep{2020ApJ...904...40R}.\ We can estimate the continuum luminosity required to produce such line strengths for simple nuclear geometries if we assume the AGN is the ionizing source.\ For an ionization parameter of $\sim700$ and a column density of $10^{23}$ cm$^{-2}$, the line luminosity we observe in the core of NGC 4945 of L$_{\rm Fe K\alpha} \sim5\times10^{38}$ ergs s$^{-1}$ requires a continuum luminosity of $\sim10^{42}$ ergs s$^{-1}$ (Kallman 1991).\footnote{Kallman. T. 1991, in Iron Line Diagnostics in X-ray Sources, ed. Treves, et al. (Berlin, Springer), 87}\ This is perfectly consistent with the observed AGN continuum luminosity of a few $\times 10^{42}$ ergs s$^{-1}$ \citep{2014ApJ...793...26P}.\ The problem is that for the extended regions we do not detect column densities this high or they would certainly be visible in the HI and CO maps.\ 

\begin{figure}[h]
\plotone{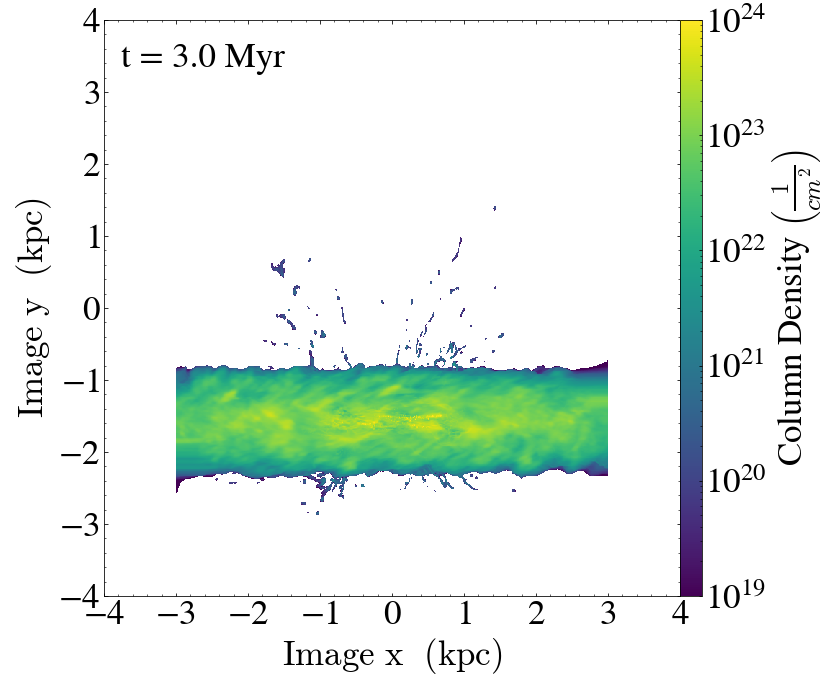}
\caption{\footnotesize{Simulation of N$_{\rm H}$ based on \citep{2022AJ....163..134T} showing the distribution of H~I column density and filamentary structure for an AGN-driven outflow event where a jet passes approximately through the disk of the galaxy at an angle of $15^{\circ}$ to $30^{\circ}$ with respect to the disk plane.\ For initial conditions, the disk gas is set to a temperature of $10^4$ K.}
\label{fig:fig11}}
\end{figure}

The hydrogen column densities in the galaxy disk and in outflowing gas are measured to be from $\sim1 - 7 \times 10^{21}$ cm$^{-2}$ up to at least a few times $10^{22}$ cm$^{-2}$ \citep{Ianjamasimanana2022, 2021ApJ...923...83B}.\ AGN-driven outflows that pass through the disk of a galaxy can drive out ISM filaments \citep{2022AJ....163..134T, 2016MNRAS.461..967M}.\ To see how a specific AGN-driven event in NGC 4945 might effect the ISM, we followed the methods outlined in \citet{2022AJ....163..134T} and examined the case where a jet event passes into the disk of the galaxy at an angle of $60^{\circ}$ to $75^{\circ}$, inclined with respect to the galaxy minor axis ($15^{\circ}$ to $30^{\circ}$ with respect to the plane).\ For a standard disk setup, an AGN-driven outflow can drive out filaments with column densities of 10$^{21-22}$ cm$^{-2}$ (Figure \ref{fig:fig11}).\ Such neutral gas filaments can show up above the galaxy plane out to at least a few kpc and so it is possible to drive gas clouds out to the distances where the Fe K$\alpha$ emission is observed.\ The question is how much fluorescence can we expect from these clouds?

In the extended regions, the  column densities are in the optically-thin regime.\ When optically thin to absorption and scattering between 6 and 7 keV, the relation between N$_{\rm H}$ and the efficiency for producing Fe K$\alpha$ depends only on the covering factor \citep{2010MNRAS.401..411Y}.\ Larger solid angles produce more line emission.\ The expected Fe K EWs should drop significantly for $10^{22}$ cm$^{-2}$ \citep{2020ApJ...904...40R} for typical geometries.\ On the other hand, the inclination of the galaxy  creates a scenario where activity will be obscured by default if it occurs behind the disk clouds. We suggest that an extended hard X-ray source could see a large solid angle of attenuating material even for lower columns along the line-of-sight.\ \citet{2010MNRAS.401..411Y} calculate the efficiency for producing Fe K$\alpha$ in a dense medium and show that this peaks between N$_{H} = 10^{23}$ cm$^{-2}$ and N$_{H} = 10^{24}$ cm$^{-2}$, whereas the efficiency drops by about a factor of 10 for N$_{H} = 10^{22}$ cm$^{-2}$.\ This allows EWs of a few hundred eV for N$_{H} = 10^{22}$ cm$^{-2}$ for a large covering factor, and so fluorescence in the ISM can produce the strength of Fe K$\alpha$ between $65''$ and $130''$ (region A3) with EW $\sim200 - 300$ eV in NGC 4945.\ It is much more difficult to explain a high EW of $\sim1000-2000$ eV.

An alternative explanation is a time delay due to a past flaring event. Large scale diffuse neutral Fe K$\alpha$ emission is detected from the molecular clouds in the center of the Milky Way \citep{1996PASJ...48..249K, 2010PASJ...62..423N, 2015MNRAS.453..172P, 2017MNRAS.468..165C, ponti_morris_clavel_terrier_goldwurm_soldi_sturm_haberl_nandra_2013}.\ The X-ray spectrum looks like a reflection spectrum with a flat power law slope and neutral Fe K$\alpha$ with a large EW of $~\sim1$ keV.\ The emission follows the cloud distribution and the required X-ray luminosity to produce this feature is about 1.5 x 10$^{39}$ erg s$^{-1}$. The strong feature likely arose from a past flaring event for Sgr A$^{\star}$ \citep{ponti_morris_clavel_terrier_goldwurm_soldi_sturm_haberl_nandra_2013}.

Since the hook spectrum in NGC 4945 is consistent with a reflected spectrum we examine whether we are seeing a light echo, similar to the reflection nebula in the Milky Way \citep{2017MNRAS.468..165C, ponti_morris_clavel_terrier_goldwurm_soldi_sturm_haberl_nandra_2013}.\ For a past flare at a maximum efficiency for Fe K production of $0.1\%$ for N$_{H} = 2\times10^{22}$ cm$^{-2}$ \citep{2010MNRAS.401..411Y} and a flare originating from the AGN, the L$_{\rm Fe K\alpha} \sim10^{38}$ ergs s$^{-1}$ implies a flare luminosity of L$_{2-10\,\rm{keV}}$ = $10^{41}$ erg s$^{-1}$ terminating $\sim16,000$ years ago.\ This estimate is based on the light travel time since the region is actively fluorescing and we do not see the corresponding nuclear activity.\ The problem with reflection of light from the AGN is that this would require a significant break in the molecular ring to allow the light to escape.\ If a flare occurred only 16,000 years ago, we should still see such a break since the dynamic timescale for the region to fall back into equilibrium is closer to 1 Myr.\ Based on the lack of ionizing UV photons \citep{2003ApJ...588..763D} that should be escaping the AGN if breaks or holes existed there is no obvious evidence for such a recent nuclear event.\ We note however that in the central regions the gas responsible
for reprocessing the nuclear AGN radiation is clumpy enough \citep{2017MNRAS.470.4039M} that nuclear activity could indeed be disrupting the dust and gas clouds.


\subsubsection{Relic activity}\label{subsubsec:relic} 

The energy contained in galactic outflows can drive their expansion for longer than the timescales that the central source is active \citep{2011MNRAS.415L...6K}.\ For velocities of $~\sim1000$ km s$^{-1}$ at a radius of $\sim1$ kpc the expansion timescale for outflowing gas is about $ 10^6$ yr \citep{2022MNRAS.516.4963I}.\ This implies that many outflows observed on kpc scales in AGN are related to episodes of past activity rather than current nuclear activity.\ The size and morphology of the Fe K$\alpha$ nebulosity in NGC 4945 coupled with a currently fully obscured AGN with no evidence for escaping ionizing UV photons bears examination as such a relic of a past event.

As we point out in \ref{subsubsec:chandra_ps}, the nuclear outflow cone of ionized gas is aligned along the galaxy minor axis.\ Shocks seem to dominate the heating of the ISM at greater than 100 pc from the nucleus \citep{2020yCat..36420166B} and optical spectra reveal large velocities of $\sim600$ km s$^{-1}$ \citep{2021ApJ...923...83B}.\ These large velocities are consistent with gas that could eventually escape into the circumgalactic medium.\ 

The hard X-ray Fe K$\alpha$ emission resides in the vicinity of two maxima of linearly polarized radio emission, which are located approximately 2.4 kpc out of the plane of the galaxy \citep{1989A&A...216...39H}.\ These two distinct polarized sources suggests a strong and ordered magnetic field at these locations, which is the type of field alignment that could be produced from a collimated outflow of relativistic particles.\ The shape of the Fe K map seems to (approximately) trace areas around the polarization maxima - possibly edge brightened regions (Figure \ref{fig:fig10}).\ If so, then the X-ray scattering may trace a relic from an ejection event or period of eruption several million years ago. This means that we could be seeing the leading part of the AGN outflow from that episode, which also coincides with the age of the starburst of 5-6 Myr \citep[][and Section \ref{subsec:starburst}]{2000A&A...357...24M}.\ We speculate that the previous AGN eruption could have triggered the starburst, with a suggested temporal sequence of: AGN ejection episode $\rightarrow$ nuclear starburst $\rightarrow$ nuclear outflow. We discuss this possible scenario in the following paragraphs below.

\begin{figure}[h]
\plotone{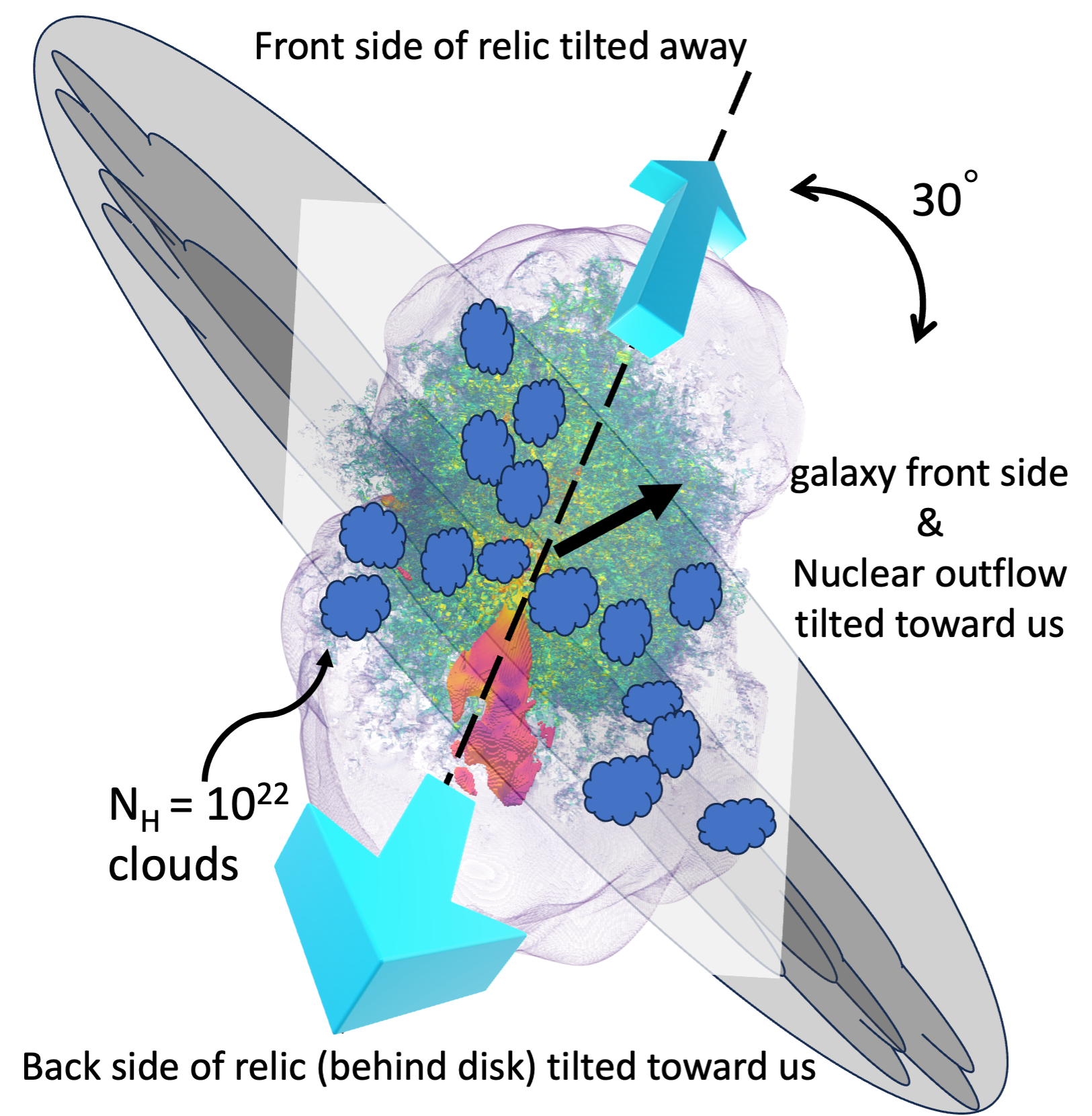}
\caption{\footnotesize{Cartoon model of AGN-driven ejection event and relic signature.\ The central image is a model from \citet{2022AJ....163..134T} with greens/yellows indicating shocked gas with high pressure (and relatively high temperature) and red/oranges indicating high velocity jet gas with v $> 0.1$ c.\ The AGN power is $10^{42}$ erg s$^{-1}$ and the jet inclination is $60^{\circ}$ with respect to the galaxy minor axis.\ The simulation is shown at 600 kyr, before the jet has turned off; it would no longer be visible.\ The large grey ellipse is the optical galaxy and the dust lanes are the darker gray ellipses. The small blue clouds represent filaments driven out of the disk by the AGN activity.\ This ejection event happens to be oriented so that it intercepts a significant fraction of the disk plane (as opposed to the nuclear outflow direction indicated by the black arrow).}}
\label{fig:fig12}
\end{figure}

Looking again at Figure \ref{fig:fig10}, the angles of the polarized radio emission coupled with the hard X-rays suggest that the original event, if AGN driven, had to occur offset from the direction of the nuclear plume that is connected with the more compact region of current star formation.\ And to accumulate enough neutral material for fluorescence to dominate at such distances, the AGN-driven outflow would need to intersect a significant part of the galaxy disk - unlike what is currently going on in the galaxy center.\ For low power AGN, \citet{2022AJ....163..134T} found that when the jet is at higher inclination angles and passing through a large column of disk gas, this can significantly affect the shape and the content of the outflow.\ The structure of the ISM within 1 kpc of the AGN also has an equally strong effect on the outflow, and the nuclear structure a few Myrs ago may have been different than it is today.\ A jet may indeed never leave (or have left) the disk, but hot bubbles of shocked ISM gas can expand outside of the disk, similar to a starburst-driven superwind.\ For a lower-power AGN of NGC 4945 luminosity, for the AGN-driven outflow to clear a path through the ISM, the activity must last for $\ge600$ kyr \citep{2022AJ....163..134T}.

We have examined the case of an AGN-driven jet and tilting the direction of the jet with respect to the minor axis of the galaxy from $60^{\circ}$ to $85^{\circ}$ (into the plane).\ We assume that the large scale hard X-ray emission is the remnant ionization from the super bubble.\ Based on the size of the apparent bubble, if we assume that the hook is the leading edge of the bubble, then the eruption happened between 2-7 million years ago, which could easily coincide with the event that sparked the starburst $\sim5$ Myr ago.\ This calculation assumes a distance of between 10-15 kpc to the leading edge of the bubble, with a bubble expansion velocity of between 2,000-5,000 km/s driven by a relativistic jet.\ The jet would obviously be faster (0.7-0.99 c) but the shock front would move much more slowly \citep[2,000-5,000 km s$^{-1}$][]{2022AJ....163..134T}.\ The shock front, even after the jet turned off, would still have residual Fe K emission for a few million years, but it would be, on a galactic scale, transitory.\ We could be seeing a transitory phenomena that may only last 1 to 3 million years and is the ``afterglow" of the short lived jet (which would last between 0.5-2.0 Myr).

Hard X-rays can also dissociate molecular clouds and in this way supress star formation.\ In Circinus, \citet{kawamuro2019} detect Fe I K$\alpha$ lines with EWs of 2.3 and  4.8 keV.\ Interestingly, they find that the molecular gas emission is faint in regions with bright Fe K emission, and they suggest that this is due to the molecular gas being efficiently dissociated by X-ray irradiation from the AGN.\ 
The parameter which determines fractional abundances of atomic and molecular gas species \citep{1996ApJ...466..561M}, is the effective ionization parameter for gas illuminated by a power-law.\ We can rewrite equation (1) from \citet{kawamuro2019} based on \citet{1996ApJ...466..561M} as L$_{44}$ = ($n_5R^2_{100}N^{\alpha}_{H,22}\xi_{eff}$)/0.1, 
where L$_{44}$, $n_5$, $R^2_{100}$, $N^{\alpha}_{H,22}$ and $\alpha$ represent the 1–100 keV luminosity in units of 
$10^{44}$ erg s$^{-1}$, the hydrogen molecular gas density in units of $10^5$ cm$^{-3}$, 
the distance in units of 100 pc, and the column density in units of $10^{22}$ cm$^{-2}$.\ The factor $\alpha$ accounts for spectral shape; a photon index of 1.7 corresponds to $\alpha\sim0.9$.

Above Log $\xi_{eff} = -3$ a large fraction of the molecular hydrogen is predicted to be dissociated into atomic hydrogen \citep{1996ApJ...466..561M}.\ For NGC 4945 we also know that the gas densities are high - as much as $10^5$ cm$^{-3}$ or higher \citep{2021ApJ...923...83B}.\ If we assume the (diffuse) X-ray source to be at a distance R=100 pc from the clouds, this predicts $\sim10^{42}$ ergs s$^{-1}$ for an ionizing luminosity.\ If we assume only R = 10 pc, this equation predicts a required 1 to 100 keV ionizing luminosity of $\sim10^{40}$ erg s$^{-1}$.\ We can't measure the 1-100 keV flux in the extended region, but for the hook we measure an extrapolated 1 to 16 keV continuum flux of $3\times10^{-13}$\,erg\,cm$^{-2}$\,s$^{-1}$, which corresponds to a ionizing luminosity of $\sim10^{39}$, not quite high enough even if we assume the clouds are embedded in the diffuse emission.\ Moving inward to locations of 1 to 2 kpc from the nucleus, the continuum luminosities are high enough $\sim10^{40-41}$ ergs s$^{-1}$ that dissociation of molecular clouds can occur.\ It's interesting that for NGC 4945, if the past AGN activity has driven a jet more or less into the plane of the galaxy, we may be at an optimum viewing angle and distance to observe directly how AGN feedback, in driving cold gas through the ISM, can act to suppress star formation.

\subsection{Future studies} 

To determine the source of the large scale X-ray emission it will be key to understand how it relates to the radio polarization and also to look for signatures of X-ray polarization.\ The polarization is geometry dependent and since there is a complex edge-on geometry in NGC 4945 with possibly multiple ejection events, looking for signatures of polarized light is crucial to understand the physics and evolution of the system.\ The illumination/reflection model predicts that the continuum emission is polarized \citep{2002MNRAS.330..817C, 2017MNRAS.468..165C} and so a simple detection of X-ray polarization could support this hypothesis.\ The polarization angle can also constrain the location of the primary source that is illuminating the clouds while the degree of polarization can constrain the line-of-sight position of the scattering clouds relative to the primary source.\ The Imaging X-ray Polarization Explorer (IXPE), which has an imaging half-power diameter of $\sim24''$, could be used to search for X-ray polarization in NGC 4945 and would enable the mapping of the extended kpc-sized regions in detail due to the proximity of NGC 4945.   

\section{Conclusions\label{sec:conclusions}}

We have analyzed XMM-Newton EPIC and Chandra data from 3 to 10 keV for the nearby starburst/AGN galaxy NGC 4945 to investigate the source of hard X-ray emission.\ We have identified a region of Fe I K$\alpha$ 6.4 keV emission that extends $\sim480$'' along the plane of NGC 4945, and to $\sim250$'' above the plane. We summarize the results of this work as follows:
\begin{itemize}
    \item We have detected neutral Fe K$\alpha$ fluorescence emission extended on scales as large as $\sim10$ kpc in NGC 4945.
    \item Fe I K$\alpha$ line emission remains strong radially out to $\sim220''$ from the galaxy center.\ The EW ranges from $\sim240$ eV to $\sim2000$ eV (and varies non-uniformly with increasing radius, see Table~\ref{tab:spectral}). 
    \item At the greatest distance we probe ($\sim130-220''$) the EW of the Fe K$\alpha$ feature is as large as or larger than at the nucleus.\ 
    \item From the EPIC pn and MOS1 radial profiles, the PSF-subtracted pn image, and spatially resolved spectra, we estimate that about 15\% of the neutral Fe K$\alpha$ emission in NGC 4945 is extended beyond distances of $\sim680$ pc from the nucleus.
    \item The entire Fe K nebulosity is angled at $\sim60^{\circ}$ with respect to the galaxy optical axis in projection and offset from the nuclear X-ray plume by $\sim30^{\circ}$ in projection.\ This alignment follows the size and angle of previously reported lobes of polarized radio emission.
    \item The luminosities, EWs and locations of the Fe K emission rule out the possibility that the Fe K nebulosity is dominated by stellar point sources.\ It is also unlikely to be related to the current star formation in the galaxy. 
    \item We have examined a scenario where the nebulosity is similar to reflection nebulae in the Milky Way where an AGN ejection event for NGC 4945  would have occurred on the order of $16,000$ years ago.\ 
    \item We have examined the case that the emission is associated with relic AGN activity - a jet event that occurred $\ge5$ Mry ago.\ This is consistent with the age of the starburst and we speculate that the AGN event may have kicked off the starburst event in this galaxy.
\end{itemize}

\begin{acknowledgments}
We acknowledge ESA for their faithful operation of the XMM-Newton satellite and the Science Operations Center. We also acknowledge
NASA for the financial support via an XMM-Newton observing grant for proposal number $22-XMMNC21-0022$ proposal \#90354.

J.\ M.\ C.\ and R.\ W.\ P.\ gratefully acknowledge support through an appointment to the NASA Postdoctoral Program at Goddard Space Flight Center, administered by ORAU through a contract with NASA. I.\ C.\ and M.\ M.\ gratefully acknowledge NASA's support under award number 80GSFC21M0002.

This paper employs a list of Chandra datasets, obtained by the Chandra X-ray Observatory, contained in the Chandra Data Collection (CDC) 220~\dataset[doi:10.25574/cdc.220]{https://doi.org/10.25574/cdc.220}

This research has made use of software provided by the \textit{Chandra} X-ray Center (CXC) in the application packages \textsc{CIAO}.

\end{acknowledgments}

%

\vspace{5mm}
\facilities{XMM-Newton (EPIC and OM), Chandra ACIS}


\software{SAS \citep{2004ASPC..314..759G},  
          CIAO \citep{2006SPIE.6270E..1VF}, 
          ftools \citep{1999ascl.soft12002B},
          matplotlib \citep{2007CSE.....9...90H},
          APLpy \citep{2012ascl.soft08017R},
          Xspec \citep{1996ASPC..101...17A},
          Mathematica \citep{Mathematica}}


\bibliography{sample631}{}

\begin{thebibliography}{}
\expandafter\ifx\csname natexlab\endcsname\relax\def\natexlab#1{#1}\fi
\providecommand{\url}[1]{\href{#1}{#1}}
\providecommand{\dodoi}[1]{doi:~\href{http://doi.org/#1}{\nolinkurl{#1}}}
\providecommand{\doeprint}[1]{\href{http://ascl.net/#1}{\nolinkurl{http://ascl.net/#1}}}
\providecommand{\doarXiv}[1]{\href{https://arxiv.org/abs/#1}{\nolinkurl{https://arxiv.org/abs/#1}}}

\bibitem[{{Agueero} \& {Carranza}(1986)}]{1986Ap&SS.121..403A}
{Agueero}, E.~L., \& {Carranza}, G.~J. 1986, \apss, 121, 403, \dodoi{10.1007/BF00653711}

\bibitem[{{Ar{\'e}valo} {et~al.}(2014){Ar{\'e}valo}, {Bauer}, {Puccetti}, {Walton}, {Koss}, {Boggs}, {Brandt}, {Brightman}, {Christensen}, {Comastri}, {Craig}, {Fuerst}, {Gandhi}, {Grefenstette}, {Hailey}, {Harrison}, {Luo}, {Madejski}, {Madsen}, {Marinucci}, {Matt}, {Saez}, {Stern}, {Stuhlinger}, {Treister}, {Urry}, \& {Zhang}}]{2014ApJ...791...81A}
{Ar{\'e}valo}, P., {Bauer}, F.~E., {Puccetti}, S., {et~al.} 2014, \apj, 791, 81, \dodoi{10.1088/0004-637X/791/2/81}

\bibitem[{{Arnaud}(1996)}]{1996ASPC..101...17A}
{Arnaud}, K.~A. 1996, in Astronomical Society of the Pacific Conference Series, Vol. 101, Astronomical Data Analysis Software and Systems V, ed. G.~H. {Jacoby} \& J.~{Barnes}, 17

\bibitem[{{Bauer} {et~al.}(2015){Bauer}, {Ar{\'e}valo}, {Walton}, {Koss}, {Puccetti}, {Gandhi}, {Stern}, {Alexander}, {Balokovi{\'c}}, {Boggs}, {Brandt}, {Brightman}, {Christensen}, {Comastri}, {Craig}, {Del Moro}, {Hailey}, {Harrison}, {Hickox}, {Luo}, {Markwardt}, {Marinucci}, {Matt}, {Rigby}, {Rivers}, {Saez}, {Treister}, {Urry}, \& {Zhang}}]{2015ApJ...812..116B}
{Bauer}, F.~E., {Ar{\'e}valo}, P., {Walton}, D.~J., {et~al.} 2015, \apj, 812, 116, \dodoi{10.1088/0004-637X/812/2/116}

\bibitem[{{Bellocchi} {et~al.}(2020){Bellocchi}, {Martin-Pintado}, {Gusten}, {Requena-Torres}, {Harris}, {van der Werf}, {Israel}, {Weiss}, {Kramer}, {Garcia-Burillo}, \& {Stutzki}}]{2020yCat..36420166B}
{Bellocchi}, E., {Martin-Pintado}, J., {Gusten}, R., {et~al.} 2020, VizieR Online Data Catalog, J/A+A/642/A166, \dodoi{10.26093/cds/vizier.36420166}

\bibitem[{{Blackburn} {et~al.}(1999){Blackburn}, {Shaw}, {Payne}, {Hayes}, \& {Heasarc}}]{1999ascl.soft12002B}
{Blackburn}, J.~K., {Shaw}, R.~A., {Payne}, H.~E., {Hayes}, J.~J.~E., \& {Heasarc}. 1999, {FTOOLS: A general package of software to manipulate FITS files}, Astrophysics Source Code Library, record ascl:9912.002.
\newblock \doeprint{9912.002}

\bibitem[{{Bolatto} {et~al.}(2021){Bolatto}, {Leroy}, {Levy}, {Meier}, {Mills}, {Thompson}, {Emig}, {Veilleux}, {Ott}, {Gorski}, {Walter}, {Lopez}, \& {Lenki{\'c}}}]{2021ApJ...923...83B}
{Bolatto}, A.~D., {Leroy}, A.~K., {Levy}, R.~C., {et~al.} 2021, \apj, 923, 83, \dodoi{10.3847/1538-4357/ac2c08}

\bibitem[{{Brandt} {et~al.}(1996){Brandt}, {Iwasawa}, \& {Reynolds}}]{1996MNRAS.281L..41B}
{Brandt}, W.~N., {Iwasawa}, K., \& {Reynolds}, C.~S. 1996, \mnras, 281, L41, \dodoi{10.1093/mnras/281.3.L41}

\bibitem[{{Chevalier} \& {Clegg}(1985)}]{1985Natur.317...44C}
{Chevalier}, R.~A., \& {Clegg}, A.~W. 1985, \nat, 317, 44, \dodoi{10.1038/317044a0}

\bibitem[{{Churazov} {et~al.}(2017){Churazov}, {Khabibullin}, {Ponti}, \& {Sunyaev}}]{2017MNRAS.468..165C}
{Churazov}, E., {Khabibullin}, I., {Ponti}, G., \& {Sunyaev}, R. 2017, \mnras, 468, 165, \dodoi{10.1093/mnras/stx443}

\bibitem[{{Churazov} {et~al.}(2002){Churazov}, {Sunyaev}, \& {Sazonov}}]{2002MNRAS.330..817C}
{Churazov}, E., {Sunyaev}, R., \& {Sazonov}, S. 2002, \mnras, 330, 817, \dodoi{10.1046/j.1365-8711.2002.05113.x}

\bibitem[{{Colbert} {et~al.}(2004){Colbert}, {Heckman}, {Ptak}, {Strickland}, \& {Weaver}}]{2004ApJ...602..231C}
{Colbert}, E. J.~M., {Heckman}, T.~M., {Ptak}, A.~F., {Strickland}, D.~K., \& {Weaver}, K.~A. 2004, \apj, 602, 231, \dodoi{10.1086/380899}

\bibitem[{{Dahlem} {et~al.}(1993){Dahlem}, {Golla}, {Whiteoak}, {Wielebinski}, {Huettemeister}, \& {Henkel}}]{dahlem1993}
{Dahlem}, M., {Golla}, G., {Whiteoak}, J.~B., {et~al.} 1993, \aap, 270, 29

\bibitem[{{Done} {et~al.}(2003){Done}, {Madejski}, {{\.Z}ycki}, \& {Greenhill}}]{2003ApJ...588..763D}
{Done}, C., {Madejski}, G.~M., {{\.Z}ycki}, P.~T., \& {Greenhill}, L.~J. 2003, \apj, 588, 763, \dodoi{10.1086/374332}

\bibitem[{{Emig} {et~al.}(2020){Emig}, {Bolatto}, {Leroy}, {Mills}, {Jim{\'e}nez Donaire}, {Tielens}, {Ginsburg}, {Gorski}, {Krieger}, {Levy}, {Meier}, {Ott}, {Rosolowsky}, {Thompson}, \& {Veilleux}}]{2020ApJ...903...50E}
{Emig}, K.~L., {Bolatto}, A.~D., {Leroy}, A.~K., {et~al.} 2020, \apj, 903, 50, \dodoi{10.3847/1538-4357/abb67d}

\bibitem[{{Fabbiano} {et~al.}(2017){Fabbiano}, {Elvis}, {Paggi}, {Karovska}, {Maksym}, {Raymond}, {Risaliti}, \& {Wang}}]{2017ApJ...842L...4F}
{Fabbiano}, G., {Elvis}, M., {Paggi}, A., {et~al.} 2017, \apjl, 842, L4, \dodoi{10.3847/2041-8213/aa7551}

\bibitem[{{Fabbiano} {et~al.}(2018{\natexlab{a}}){Fabbiano}, {Paggi}, {Karovska}, {Elvis}, {Maksym}, {Risaliti}, \& {Wang}}]{2018ApJ...855..131F}
{Fabbiano}, G., {Paggi}, A., {Karovska}, M., {et~al.} 2018{\natexlab{a}}, \apj, 855, 131, \dodoi{10.3847/1538-4357/aab1f4}

\bibitem[{{Fabbiano} {et~al.}(2018{\natexlab{b}}){Fabbiano}, {Paggi}, {Karovska}, {Elvis}, {Maksym}, \& {Wang}}]{2018ApJ...865...83F}
---. 2018{\natexlab{b}}, \apj, 865, 83, \dodoi{10.3847/1538-4357/aadc5d}

\bibitem[{{Fabbiano} {et~al.}(2019){Fabbiano}, {Siemiginowska}, {Paggi}, {Elvis}, {Volonteri}, {Mayer}, {Karovska}, {Maksym}, {Risaliti}, \& {Wang}}]{2019ApJ...870...69F}
{Fabbiano}, G., {Siemiginowska}, A., {Paggi}, A., {et~al.} 2019, \apj, 870, 69, \dodoi{10.3847/1538-4357/aaf0a4}

\bibitem[{{Fruscione} {et~al.}(2006){Fruscione}, {McDowell}, {Allen}, {Brickhouse}, {Burke}, {Davis}, {Durham}, {Elvis}, {Galle}, {Harris}, {Huenemoerder}, {Houck}, {Ishibashi}, {Karovska}, {Nicastro}, {Noble}, {Nowak}, {Primini}, {Siemiginowska}, {Smith}, \& {Wise}}]{2006SPIE.6270E..1VF}
{Fruscione}, A., {McDowell}, J.~C., {Allen}, G.~E., {et~al.} 2006, in Society of Photo-Optical Instrumentation Engineers (SPIE) Conference Series, Vol. 6270, Society of Photo-Optical Instrumentation Engineers (SPIE) Conference Series, ed. D.~R. {Silva} \& R.~E. {Doxsey}, 62701V, \dodoi{10.1117/12.671760}

\bibitem[{{Gabriel} {et~al.}(2004){Gabriel}, {Denby}, {Fyfe}, {Hoar}, {Ibarra}, {Ojero}, {Osborne}, {Saxton}, {Lammers}, \& {Vacanti}}]{2004ASPC..314..759G}
{Gabriel}, C., {Denby}, M., {Fyfe}, D.~J., {et~al.} 2004, in Astronomical Society of the Pacific Conference Series, Vol. 314, Astronomical Data Analysis Software and Systems (ADASS) XIII, ed. F.~{Ochsenbein}, M.~G. {Allen}, \& D.~{Egret}, 759

\bibitem[{{Gallimore} {et~al.}(2016){Gallimore}, {Elitzur}, {Maiolino}, {Marconi}, {O'Dea}, {Lutz}, {Baum}, {Nikutta}, {Impellizzeri}, {Davies}, {Kimball}, \& {Sani}}]{2016ApJ...829L...7G}
{Gallimore}, J.~F., {Elitzur}, M., {Maiolino}, R., {et~al.} 2016, \apjl, 829, L7, \dodoi{10.3847/2041-8205/829/1/L7}

\bibitem[{{Garc{\'\i}a-Burillo} {et~al.}(2016){Garc{\'\i}a-Burillo}, {Combes}, {Ramos Almeida}, {Usero}, {Krips}, {Alonso-Herrero}, {Aalto}, {Casasola}, {Hunt}, {Mart{\'\i}n}, {Viti}, {Colina}, {Costagliola}, {Eckart}, {Fuente}, {Henkel}, {M{\'a}rquez}, {Neri}, {Schinnerer}, {Tacconi}, \& {van der Werf}}]{2016ApJ...823L..12G}
{Garc{\'\i}a-Burillo}, S., {Combes}, F., {Ramos Almeida}, C., {et~al.} 2016, \apjl, 823, L12, \dodoi{10.3847/2041-8205/823/1/L12}

\bibitem[{{Gehrels}(1986)}]{1986ApJ...303..336G}
{Gehrels}, N. 1986, \apj, 303, 336, \dodoi{10.1086/164079}

\bibitem[{{Greenhill} {et~al.}(1997){Greenhill}, {Moran}, \& {Herrnstein}}]{1997ApJ...481L..23G}
{Greenhill}, L.~J., {Moran}, J.~M., \& {Herrnstein}, J.~R. 1997, \apjl, 481, L23, \dodoi{10.1086/310643}

\bibitem[{{Guainazzi} {et~al.}(2000){Guainazzi}, {Matt}, {Brandt}, {Antonelli}, {Barr}, \& {Bassani}}]{2000A&A...356..463G}
{Guainazzi}, M., {Matt}, G., {Brandt}, W.~N., {et~al.} 2000, \aap, 356, 463, \dodoi{10.48550/arXiv.astro-ph/0001528}

\bibitem[{{Harnett} {et~al.}(1989){Harnett}, {Haynes}, {Klein}, \& {Wielebinski}}]{1989A&A...216...39H}
{Harnett}, J.~I., {Haynes}, R.~F., {Klein}, U., \& {Wielebinski}, R. 1989, \aap, 216, 39

\bibitem[{{Heckman} {et~al.}(1990){Heckman}, {Armus}, \& {Miley}}]{1990ApJS...74..833H}
{Heckman}, T.~M., {Armus}, L., \& {Miley}, G.~K. 1990, \apjs, 74, 833, \dodoi{10.1086/191522}

\bibitem[{Heckman \& Best(2014)}]{Heckman_2014}
Heckman, T.~M., \& Best, P.~N. 2014, Annual Review of Astronomy and Astrophysics, 52, 589, \dodoi{10.1146/annurev-astro-081913-035722}

\bibitem[{{Heckman} {et~al.}(2004){Heckman}, {Kauffmann}, {Brinchmann}, {Charlot}, {Tremonti}, \& {White}}]{2004ApJ...613..109H}
{Heckman}, T.~M., {Kauffmann}, G., {Brinchmann}, J., {et~al.} 2004, \apj, 613, 109, \dodoi{10.1086/422872}

\bibitem[{{Hunter}(2007)}]{2007CSE.....9...90H}
{Hunter}, J.~D. 2007, Computing in Science and Engineering, 9, 90, \dodoi{10.1109/MCSE.2007.55}

\bibitem[{Huo {et~al.}(2015)Huo, Li, Li, \& Zhou}]{Huo_2015}
Huo, Z.-X., Li, Y.-M., Li, X.-B., \& Zhou, J.-F. 2015, Research in Astronomy and Astrophysics, 15, 1905, \dodoi{10.1088/1674-4527/15/11/012}

\bibitem[{{Ianjamasimanana} {et~al.}(2022){Ianjamasimanana}, {Koribalski}, {J{\'o}zsa}, {Kamphuis}, {de Blok}, {Kleiner}, {Namumba}, {Carignan}, {Dettmar}, {Serra}, {Smirnov}, {Thorat}, {Hugo}, {Ramaila}, {Maina}, {Maccagni}, {Makhathini}, {Andati}, {Moln{\'a}r}, {Perkins}, {Loi}, {Ramatsoku}, \& {Atemkeng}}]{Ianjamasimanana2022}
{Ianjamasimanana}, R., {Koribalski}, B.~S., {J{\'o}zsa}, G. I.~G., {et~al.} 2022, \mnras, 513, 2019, \dodoi{10.1093/mnras/stac936}

\bibitem[{{Imanishi} {et~al.}(2016){Imanishi}, {Nakanishi}, \& {Izumi}}]{2016ApJ...822L..10I}
{Imanishi}, M., {Nakanishi}, K., \& {Izumi}, T. 2016, \apjl, 822, L10, \dodoi{10.3847/2041-8205/822/1/L10}

\bibitem[{{Ishibashi} \& {Fabian}(2022)}]{2022MNRAS.516.4963I}
{Ishibashi}, W., \& {Fabian}, A.~C. 2022, \mnras, 516, 4963, \dodoi{10.1093/mnras/stac2614}

\bibitem[{{Jones} {et~al.}(2020){Jones}, {Fabbiano}, {Elvis}, {Paggi}, {Karovska}, {Maksym}, {Siemiginowska}, \& {Raymond}}]{2020ApJ...891..133J}
{Jones}, M.~L., {Fabbiano}, G., {Elvis}, M., {et~al.} 2020, \apj, 891, 133, \dodoi{10.3847/1538-4357/ab76c8}

\bibitem[{{Karachentsev} \& {Makarov}(1996)}]{1996AJ....111..794K}
{Karachentsev}, I.~D., \& {Makarov}, D.~A. 1996, \aj, 111, 794, \dodoi{10.1086/117825}

\bibitem[{{Kawamuro} {et~al.}(2019){Kawamuro}, {Izumi}, \& {Imanishi}}]{kawamuro2019}
{Kawamuro}, T., {Izumi}, T., \& {Imanishi}, M. 2019, \pasj, 71, 68, \dodoi{10.1093/pasj/psz045}

\bibitem[{{King} {et~al.}(2011){King}, {Zubovas}, \& {Power}}]{2011MNRAS.415L...6K}
{King}, A.~R., {Zubovas}, K., \& {Power}, C. 2011, \mnras, 415, L6, \dodoi{10.1111/j.1745-3933.2011.01067.x}

\bibitem[{{Koyama} {et~al.}(1996){Koyama}, {Maeda}, {Sonobe}, {Takeshima}, {Tanaka}, \& {Yamauchi}}]{1996PASJ...48..249K}
{Koyama}, K., {Maeda}, Y., {Sonobe}, T., {et~al.} 1996, \pasj, 48, 249, \dodoi{10.1093/pasj/48.2.249}

\bibitem[{{Levenson} {et~al.}(2002){Levenson}, {Krolik}, {{\.Z}ycki}, {Heckman}, {Weaver}, {Awaki}, \& {Terashima}}]{2002ApJ...573L..81L}
{Levenson}, N.~A., {Krolik}, J.~H., {{\.Z}ycki}, P.~T., {et~al.} 2002, \apjl, 573, L81, \dodoi{10.1086/342092}

\bibitem[{{Lin} {et~al.}(2012){Lin}, {Webb}, \& {Barret}}]{2012ApJ...756...27L}
{Lin}, D., {Webb}, N.~A., \& {Barret}, D. 2012, \apj, 756, 27, \dodoi{10.1088/0004-637X/756/1/27}

\bibitem[{{Liu} {et~al.}(2014){Liu}, {Gou}, {Yuan}, \& {Mao}}]{liu2014}
{Liu}, J., {Gou}, L., {Yuan}, W., \& {Mao}, S. 2014, \mnras, 437, L76, \dodoi{10.1093/mnrasl/slt145}

\bibitem[{{Lucero} {et~al.}(2015){Lucero}, {Carignan}, {Elson}, {Randriamampandry}, {Jarrett}, {Oosterloo}, \& {Heald}}]{2015MNRAS.450.3935L}
{Lucero}, D.~M., {Carignan}, C., {Elson}, E.~C., {et~al.} 2015, \mnras, 450, 3935, \dodoi{10.1093/mnras/stv856}

\bibitem[{{Lumb} {et~al.}(2002){Lumb}, {Warwick}, {Page}, \& {De Luca}}]{2002A&A...389...93L}
{Lumb}, D.~H., {Warwick}, R.~S., {Page}, M., \& {De Luca}, A. 2002, \aap, 389, 93, \dodoi{10.1051/0004-6361:20020531}

\bibitem[{{Lyu} \& {Rieke}(2021)}]{2021ApJ...912..126L}
{Lyu}, J., \& {Rieke}, G.~H. 2021, \apj, 912, 126, \dodoi{10.3847/1538-4357/abee14}

\bibitem[{{Ma} {et~al.}(2020){Ma}, {Elvis}, {Fabbiano}, {Balokovi{\'c}}, {Maksym}, {Jones}, \& {Risaliti}}]{2020ApJ...900..164M}
{Ma}, J., {Elvis}, M., {Fabbiano}, G., {et~al.} 2020, \apj, 900, 164, \dodoi{10.3847/1538-4357/abacbe}

\bibitem[{{Maloney} {et~al.}(1996){Maloney}, {Hollenbach}, \& {Tielens}}]{1996ApJ...466..561M}
{Maloney}, P.~R., {Hollenbach}, D.~J., \& {Tielens}, A.~G.~G.~M. 1996, \apj, 466, 561, \dodoi{10.1086/177532}

\bibitem[{{Marconi} {et~al.}(2000){Marconi}, {Oliva}, {van der Werf}, {Maiolino}, {Schreier}, {Macchetto}, \& {Moorwood}}]{2000A&A...357...24M}
{Marconi}, A., {Oliva}, E., {van der Werf}, P.~P., {et~al.} 2000, \aap, 357, 24, \dodoi{10.48550/arXiv.astro-ph/0002244}

\bibitem[{{Marinucci} {et~al.}(2017){Marinucci}, {Bianchi}, {Fabbiano}, {Matt}, {Risaliti}, {Nardini}, \& {Wang}}]{2017MNRAS.470.4039M}
{Marinucci}, A., {Bianchi}, S., {Fabbiano}, G., {et~al.} 2017, \mnras, 470, 4039, \dodoi{10.1093/mnras/stx1551}

\bibitem[{{Marinucci} {et~al.}(2012){Marinucci}, {Risaliti}, {Wang}, {Nardini}, {Elvis}, {Fabbiano}, {Bianchi}, \& {Matt}}]{2012MNRAS.423L...6M}
{Marinucci}, A., {Risaliti}, G., {Wang}, J., {et~al.} 2012, \mnras, 423, L6, \dodoi{10.1111/j.1745-3933.2012.01232.x}

\bibitem[{{Mitsuishi} {et~al.}(2011){Mitsuishi}, {Yamasaki}, \& {Takei}}]{2011ApJ...742L..31M}
{Mitsuishi}, I., {Yamasaki}, N.~Y., \& {Takei}, Y. 2011, \apjl, 742, L31, \dodoi{10.1088/2041-8205/742/2/L31}

\bibitem[{{Moorwood} {et~al.}(1996){Moorwood}, {van der Werf}, {Kotilainen}, {Marconi}, \& {Oliva}}]{1996A&A...308L...1M}
{Moorwood}, A.~F.~M., {van der Werf}, P.~P., {Kotilainen}, J.~K., {Marconi}, A., \& {Oliva}, E. 1996, \aap, 308, L1

\bibitem[{{Mueller} {et~al.}(2004){Mueller}, {Madejski}, {Done}, \& {Zycki}}]{2004AIPC..714..190M}
{Mueller}, M., {Madejski}, G., {Done}, C., \& {Zycki}, P. 2004, in American Institute of Physics Conference Series, Vol. 714, X-ray Timing 2003: Rossi and Beyond, ed. P.~{Kaaret}, F.~K. {Lamb}, \& J.~H. {Swank}, 190--193, \dodoi{10.1063/1.1781025}

\bibitem[{{Mukherjee} {et~al.}(2016){Mukherjee}, {Bicknell}, {Sutherland}, \& {Wagner}}]{2016MNRAS.461..967M}
{Mukherjee}, D., {Bicknell}, G.~V., {Sutherland}, R., \& {Wagner}, A. 2016, \mnras, 461, 967, \dodoi{10.1093/mnras/stw1368}

\bibitem[{{Murphy} \& {Yaqoob}(2009)}]{2009MNRAS.397.1549M}
{Murphy}, K.~D., \& {Yaqoob}, T. 2009, \mnras, 397, 1549, \dodoi{10.1111/j.1365-2966.2009.15025.x}

\bibitem[{{Nakai}(1989)}]{1989PASJ...41.1107N}
{Nakai}, N. 1989, \pasj, 41, 1107

\bibitem[{Nir {et~al.}(2018)Nir, Zackay, \& Ofek}]{Nir_2018}
Nir, G., Zackay, B., \& Ofek, E.~O. 2018, The Astronomical Journal, 156, 229, \dodoi{10.3847/1538-3881/aaddff}

\bibitem[{{Nobukawa} {et~al.}(2010){Nobukawa}, {Koyama}, {Tsuru}, {Ryu}, \& {Tatischeff}}]{2010PASJ...62..423N}
{Nobukawa}, M., {Koyama}, K., {Tsuru}, T.~G., {Ryu}, S.~G., \& {Tatischeff}, V. 2010, \pasj, 62, 423, \dodoi{10.1093/pasj/62.2.423}

\bibitem[{{Ott} {et~al.}(2001){Ott}, {Whiteoak}, {Henkel}, \& {Wielebinski}}]{2001A&A...372..463O}
{Ott}, M., {Whiteoak}, J.~B., {Henkel}, C., \& {Wielebinski}, R. 2001, \aap, 372, 463, \dodoi{10.1051/0004-6361:20010505}

\bibitem[{{Peterson}(1980)}]{peterson1980}
{Peterson}, C.~J. 1980, \pasp, 92, 397, \dodoi{10.1086/130685}

\bibitem[{{Pier} \& {Krolik}(1992)}]{1992ApJ...401...99P}
{Pier}, E.~A., \& {Krolik}, J.~H. 1992, \apj, 401, 99, \dodoi{10.1086/172042}

\bibitem[{Ponti {et~al.}(2013)Ponti, Morris, Clavel, Terrier, Goldwurm, Soldi, Sturm, Haberl, \& Nandra}]{ponti_morris_clavel_terrier_goldwurm_soldi_sturm_haberl_nandra_2013}
Ponti, G., Morris, M.~R., Clavel, M., {et~al.} 2013, Proceedings of the International Astronomical Union, 9, 333–343, \dodoi{10.1017/S174392131400088X}

\bibitem[{{Ponti} {et~al.}(2015){Ponti}, {Morris}, {Terrier}, {Haberl}, {Sturm}, {Clavel}, {Soldi}, {Goldwurm}, {Predehl}, {Nandra}, {B{\'e}langer}, {Warwick}, \& {Tatischeff}}]{2015MNRAS.453..172P}
{Ponti}, G., {Morris}, M.~R., {Terrier}, R., {et~al.} 2015, \mnras, 453, 172, \dodoi{10.1093/mnras/stv1331}

\bibitem[{{Puccetti} {et~al.}(2014){Puccetti}, {Comastri}, {Fiore}, {Ar{\'e}valo}, {Risaliti}, {Bauer}, {Brandt}, {Stern}, {Harrison}, {Alexander}, {Boggs}, {Christensen}, {Craig}, {Gandhi}, {Hailey}, {Koss}, {Lansbury}, {Luo}, {Madejski}, {Matt}, {Walton}, \& {Zhang}}]{2014ApJ...793...26P}
{Puccetti}, S., {Comastri}, A., {Fiore}, F., {et~al.} 2014, \apj, 793, 26, \dodoi{10.1088/0004-637X/793/1/26}

\bibitem[{{Rahin} \& {Behar}(2020)}]{2020ApJ...904...40R}
{Rahin}, R., \& {Behar}, E. 2020, \apj, 904, 40, \dodoi{10.3847/1538-4357/abbb2f}

\bibitem[{{Robitaille} \& {Bressert}(2012)}]{2012ascl.soft08017R}
{Robitaille}, T., \& {Bressert}, E. 2012, {APLpy: Astronomical Plotting Library in Python}, Astrophysics Source Code Library, record ascl:1208.017.
\newblock \doeprint{1208.017}

\bibitem[{{Schurch} {et~al.}(2002){Schurch}, {Roberts}, \& {Warwick}}]{2002MNRAS.335..241S}
{Schurch}, N.~J., {Roberts}, T.~P., \& {Warwick}, R.~S. 2002, \mnras, 335, 241, \dodoi{10.1046/j.1365-8711.2002.05585.x}

\bibitem[{{Spoon} {et~al.}(2000){Spoon}, {Koornneef}, {Moorwood}, {Lutz}, \& {Tielens}}]{2000A&A...357..898S}
{Spoon}, H.~W.~W., {Koornneef}, J., {Moorwood}, A.~F.~M., {Lutz}, D., \& {Tielens}, A.~G.~G.~M. 2000, \aap, 357, 898, \dodoi{10.48550/arXiv.astro-ph/0003457}

\bibitem[{{Strickland} \& {Heckman}(2007)}]{2007ApJ...658..258S}
{Strickland}, D.~K., \& {Heckman}, T.~M. 2007, \apj, 658, 258, \dodoi{10.1086/511174}

\bibitem[{{Swartz} {et~al.}(2004){Swartz}, {Ghosh}, {Tennant}, \& {Wu}}]{swartz2004}
{Swartz}, D.~A., {Ghosh}, K.~K., {Tennant}, A.~F., \& {Wu}, K. 2004, \apjs, 154, 519, \dodoi{10.1086/422842}

\bibitem[{{Tanimoto} {et~al.}(2018){Tanimoto}, {Ueda}, {Kawamuro}, {Ricci}, {Awaki}, \& {Terashima}}]{2018ApJ...853..146T}
{Tanimoto}, A., {Ueda}, Y., {Kawamuro}, T., {et~al.} 2018, \apj, 853, 146, \dodoi{10.3847/1538-4357/aaa47c}

\bibitem[{{Tanner} \& {Weaver}(2022)}]{2022AJ....163..134T}
{Tanner}, R., \& {Weaver}, K.~A. 2022, \aj, 163, 134, \dodoi{10.3847/1538-3881/ac4d23}

\bibitem[{{Torrej{\'o}n} {et~al.}(2010){Torrej{\'o}n}, {Schulz}, {Nowak}, \& {Kallman}}]{2010ApJ...715..947T}
{Torrej{\'o}n}, J.~M., {Schulz}, N.~S., {Nowak}, M.~A., \& {Kallman}, T.~R. 2010, \apj, 715, 947, \dodoi{10.1088/0004-637X/715/2/947}

\bibitem[{{Veilleux} {et~al.}(2005){Veilleux}, {Cecil}, \& {Bland-Hawthorn}}]{2005ARA&A..43..769V}
{Veilleux}, S., {Cecil}, G., \& {Bland-Hawthorn}, J. 2005, \araa, 43, 769, \dodoi{10.1146/annurev.astro.43.072103.150610}

\bibitem[{Venturi {et~al.}(2017)Venturi, Marconi, Mingozzi, Carniani, Cresci, Risaliti, \& Mannucci}]{10.3389/fspas.2017.00046}
Venturi, G., Marconi, A., Mingozzi, M., {et~al.} 2017, Frontiers in Astronomy and Space Sciences, 4, \dodoi{10.3389/fspas.2017.00046}

\bibitem[{{Vulic} {et~al.}(2018){Vulic}, {Hornschemeier}, {Wik}, {Yukita}, {Zezas}, {Ptak}, {Lehmer}, {Antoniou}, {Maccarone}, {Williams}, \& {Fornasini}}]{2018ApJ...864..150V}
{Vulic}, N., {Hornschemeier}, A.~E., {Wik}, D.~R., {et~al.} 2018, \apj, 864, 150, \dodoi{10.3847/1538-4357/aad500}

\bibitem[{{Wang} {et~al.}(2014){Wang}, {Nardini}, {Fabbiano}, {Karovska}, {Elvis}, {Pellegrini}, {Max}, {Risaliti}, {U}, \& {Zezas}}]{2014ApJ...781...55W}
{Wang}, J., {Nardini}, E., {Fabbiano}, G., {et~al.} 2014, \apj, 781, 55, \dodoi{10.1088/0004-637X/781/1/55}

\bibitem[{{Wang} {et~al.}(2016{\natexlab{a}}){Wang}, {Liu}, {Qiu}, {Bai}, {Yang}, {Guo}, \& {Zhang}}]{2016ApJS..224...40W}
{Wang}, S., {Liu}, J., {Qiu}, Y., {et~al.} 2016{\natexlab{a}}, \apjs, 224, 40, \dodoi{10.3847/0067-0049/224/2/40}

\bibitem[{{Wang} {et~al.}(2016{\natexlab{b}}){Wang}, {Liu}, {Qiu}, {Bai}, {Yang}, {Guo}, \& {Zhang}}]{wang2016}
---. 2016{\natexlab{b}}, \apjs, 224, 40, \dodoi{10.3847/0067-0049/224/2/40}

\bibitem[{Wolfram~Research(2023)}]{Mathematica}
Wolfram~Research, I. 2023, {Mathematica, {V}ersion 13.3}.
\newblock \url{https://www.wolfram.com/mathematica}

\bibitem[{{Yaqoob} {et~al.}(2010){Yaqoob}, {Murphy}, {Miller}, \& {Turner}}]{2010MNRAS.401..411Y}
{Yaqoob}, T., {Murphy}, K.~D., {Miller}, L., \& {Turner}, T.~J. 2010, \mnras, 401, 411, \dodoi{10.1111/j.1365-2966.2009.15657.x}

\end{thebibliography}
\bibliographystyle{aasjournal}



\end{document}